\documentclass[12pt]{iopart}

\usepackage{iopams}  
\usepackage{graphicx}
\usepackage{ulem}
\usepackage{color}
\usepackage{colortbl}
\usepackage{caption}
\usepackage{subfig}
\usepackage{comment}

\usepackage{dcolumn}% Align table columns on decimal point
\usepackage{bm}% bold math
\usepackage[titletoc,title]{appendix}
\usepackage[utf8]{inputenc}
\usepackage{comment}

\newcommand{\beq}{\begin{eqnarray}}
\newcommand{\eeq}{\end{eqnarray}}
\newcommand{\bit}{\begin{itemize}}
\newcommand{\eit}{\end{itemize}}

\newcommand\ex{\textrm{e}}

\newcommand{\derivative}[2]{\frac{\mathrm{d}#2}{\mathrm{d}#1}}

\newcommand{\pderiv}[2]{\frac{\partial#2}{\partial#1}}
\newcommand{\funcderiv}[2]{\frac{\delta#2}{\delta#1}}
\newcommand{\intd}[1]{\int\mathrm{d}#1}
\newcommand{\inte}[3]{\int_{#2}^{#3}\mathrm{d}#1}
\newcommand{\mean}[1]{\langle{#1}\rangle}

\begin{document}

\title[Path Integral for Active
Particles in Harmonic
Potentials]{Stochastic Path Integral for the Active Brownian Particle in a Harmonic Potential}

\author{Carsten Littek$^{1}$, Mike Brandt$^{1}$ and Falko Ziebert$^{1}$ }

\address{$^1$Institute for Theoretical Physics, Heidelberg University, 69120 Heidelberg, Germany}
\ead{f.ziebert@thphys.uni-heidelberg.de}
\vspace{10pt}

\begin{abstract}
In this work we develop and apply a path integral formulation
for the microscopic degrees of freedom obeying stochastic differential equations
to an active Brownian particle (ABP) trapped in a harmonic potential. 
The formalism allows to derive exact analytic expressions for the 
time-dependent moments, like the mean position and the mean square displacement,
including full dependence on initial conditions. In addition, the
probability distribution of the particle's position can be evaluated systematically
as a series expansion in the propulsion speed. Compared to previous methods
relying on eigenfunction expansions of the equivalent Fokker-Planck equation, 
our method is easier to generalize to more complex situations:
it does not rely on eigenfunctions but on a reference state that can be solved analytically,
which in our case is the passive Brownian particle in a harmonic potential. 
We exemplify this versatility by also briefly treating an ABP with an active 
torque (Brownian circle swimmer, BCS) in a harmonic potential.
\end{abstract}

\section{Introduction}

Active Brownian particles (ABPs) \cite{Romanczuk} are an archetypical model class for objects
that are self-propelled, i.e.~move directionally using internal or external energy sources,
while at the same time being subject to substantial noise, either due to their small size
or because of intrinsic noise due to the motility machinery.
Important examples include swimming bacteria \cite{Bergbook,WadhwaBerg,BetaStark}, 
crawling cells \cite{FletcherTheriot,Hakim_review}
or gliding microorganisms \cite{WadhwaBerg,Leon_prep}
on the biology side, as well as active Janus colloids \cite{Howse_Ramin07}
or colloidal rollers \cite{BartoloQuincke,IgorMagnetic} on the man-made side, 
see also Ref.~\cite{RevModPhys.88.045006}
for a review.

To describe active particles, early on Langevin-type equations for
the position and the angle of direction have been proposed, 
that describe quantities extracted from experiments
such as the mean square displacement 
very well \cite{Howse_Ramin07,RevModPhys.88.045006}.
Later on, active particles in a potential were discussed theoretically \cite{Szamel}.
Experimentally, however, the realization of a trap is not easy, but recently substantial progress
has been achieved \cite{Bechinger_trap,exp_multistable}.
Other interesting topics include active particles with an additional torque
leading to circle swimmers \cite{circle_simmer} 
(BCS, Brownian circle swimmer) 
as also realized experimentally \cite{Bechinger_circle,circle_swim_newer}, 
the effect of inertia \cite{Loewen} and, of course, collective effects 
including flocking \cite{BartoloQuincke,IgorMagnetic}
as predicted by the famous Vicsek model \cite{Vicsek,Ginelli}.

Simulations of ABPs are straightforward on the level of the Langevin equations. 
Analytically, the single free particle problem is easy, but 
the problem of a single particle in a harmonic potential -- mimicking, e.g. an optical trap -- 
is unexpectedly hard. 
The reason is that for the free particle the orientational problem can be solved separately,
while in the trapped case the orientation vector induces a nonlinear coupling,
although the force derived from the potential is linear. 
After some progress in limiting cases \cite{Schehr19,Malakar20}, the problem 
was solved (semi-)analytically recently            
\cite{PhysRevLett.129.158001} in form of an infinite expansion in special functions, 
as well as by perturbation theory \cite{Pruessner24}.
These approaches involve the Fokker-Planck (or Smoluchowski) equations and aim to
diagonalize the respective operator (or transform it to a workable triangular form). 
Depending on the formalism, this leads to complicated eigenfunctions 
involving generalized Laguerre polynomials or Matthieu functions.  

While  insightful and important, this diagonalization approach is hard to 
generalize to more complex problems, like e.g.~the circle swimmer or systems
with particle interactions like the Vicsek model, which 
also has been studied in a harmonic trap recently \cite{RafaLuis}. 
Therefore we here propose a path-integral approach, based on the  
Martin-Siggia-Rose-Janssen-DeDominicis formalism, as suggested by  
a work by Mazenko \cite{PhysRevE.81.061102}. 
Within this formalism, the problem of a passive particle in a harmonic potential
(equivalent to the Ornstein-Uhlenbeck process) can be solved completely analytically.
Using the passive particle -- in the harmonic trap -- as the reference state, 
the activity can then be treated by systematic perturbation theory. 
Importantly, as was the case in  Ref.~\cite{PhysRevLett.129.158001},
the moments/correlation functions terminate at finite order in the expansion,
leading to {\it exact} analytical results for them, and the full probability distribution 
is given as a series expression. 

This work is organized as follows: in section \ref{secPathInt} we review
the general formalism of path integrals for stochastic differential equations. 
In section \ref{secActive} we formulate the problem for the 2D active particle 
in a harmonic potential.
We then solve the passive particle in the potential analytically 
-- yielding the well-known Ornstein-Uhlenbeck result -- for later use as the reference state.
Section \ref{secResults} presents how standard perturbation theory can be used
to achieve analytical results for the mean position, the MSD and the distribution function. 
In section \ref{secCircle} we demonstrate the versatility of the approach 
by briefly discussing the Brownian circle swimmer in a harmonic potential
and end in section \ref{secDisc} with a discussion and outlook.

\section{Path Integral for Stochastic Differential Equations}
\label{secPathInt}
\subsection{Equations of Motion}
We consider a system characterized by its degrees of freedom $R_\mu(t)$, such as the
position $q_\mu\in\mathbb{R}^d$ 
of a particle 
in $d$-dimensional space. The equations of motion are given by
\beq
    \mathrm{EoM}_\mu(t):=\derivative{t}{R_\mu} - F_\mu[R(t)] - \sigma_{\mu\nu}[R(t)]\eta_\nu(t)\equiv 0.\label{eq:stochastic-eom}
\eeq
Here, the function $\eta_\mu(t)$ is a random process equivalent to Gaussian white noise with
\beq
\left<\eta_\mu(t)\right>=0,\qquad\left<\eta_\mu(t)\eta_\nu(t')\right>=\delta_{\mu\nu}\delta(t-t').\label{eq:noisecorr}
\eeq
Thus, Eq.~(\ref{eq:stochastic-eom}) is a stochastic differential equation of first order, 
where Greek indices label the degrees of freedom. The functions $F_\mu$ and $\sigma_{\mu\nu}$ 
are smooth functions of $R_\mu(t)$. %The function 
$F_\mu[R(t)]$ describes the force acting on the particle, 
such as an external potential force. For example, the particle could be trapped in a harmonic potential 
$F_\mu=-\alpha\,q_\mu$ with Hooke's constant $\alpha>0$ and particle position $q_\mu(t)$. 
%The function 
$\sigma_{\mu\nu}[R(t)]$ is the variance or diffusion function. 
For $\sigma=\mathrm{const.}$~the noise is called additive, while noise is called multiplicative if the diffusion
is state-dependent. In the case of Smoluchowski dynamics the noise is additive, 
since the diffusion function is constant and given by
\beq
    \sigma_{\mu\nu}[R(t)] = \sqrt{2k_\mathrm{B}T}\,\tilde{D}_{\mu}\delta_{\mu\nu}=\mathrm{const.},
\eeq
as demanded by the fluctuation-dissipation theorem.
Note that each degree of freedom diffuses independently 
and the diffusion coefficient $\tilde{D}_{\mu}$ may vary for different degrees of freedom.
For instance, if one degree of freedom is the orientation angle, the rotational diffusion
is different from the translational one.
The entries of the diffusion matrix represent the inverse timescales of the different diffusion processes. 
For the rest of this work we assume additive noise only. 
A path integral for multiplicative noise was presented in \cite{PhysRevE.91.042103,PhysRevE.76.011123}.

\subsection{The Generating Functional}
It is customary to investigate such systems by means of the particle distribution function
$P(R_\mu,t)=\left<\delta(R_\mu-R_\mu(t))\right>$, which then obeys a Fokker-Planck-equation. 
Here we choose a description by a path integral formulated in terms of the microscopic degrees 
of freedom $R_\mu(t)$ along the lines of \cite{PhysRevE.81.061102,DasMazenko}. 
The key advantage lies in the microscopic degrees of freedom being Gaussian random variables 
while the particle distribution function typically is not.

In the following we sketch the derivation of the generating functional for a stochastic differential equation with additive Gaussian noise. A more detailed calculation is presented in \ref{app1}. We refer the interested reader to the rigorous derivation in \cite{PhysRevE.91.042103}, where also multiplicative noise is treated.

\subsubsection{Transition Probability.}
The generating functional is constructed from a path integral by weighting all paths with the probability of the system to transition from an initial state $R_0$ to a later state $R(t)$. The total probability is
\beq
    P[R(t)] = \int\mathcal{D}\eta\,\intd{R_0\,}~P\left[R(t)|R_0,\eta(t)\right]P(R_0)P[\eta(t)].
\eeq
Since we are dealing with a stochastic differential equation, 
the transition probability is conditioned on the initial values $R_0$ and on the particular realisation of the 
random processes $\eta_\mu(t)$. Only paths that solve the equations of motion (\ref{eq:stochastic-eom}) 
can contribute to the path integral. Thus, this conditional probability may be expressed by
\beq
    P[R(t)|R_0,\eta(t)] = \delta[\mathrm{EoM}_\mu(t)]\,\mathrm{det}\left[\funcderiv{R_\nu(t')}{\mathrm{EoM}_\mu(t)}\right].
    \label{eq:transition-probability}
\eeq
This form simply expresses the constraints of the dynamical equations for the system under consideration.

\subsubsection{Martin-Siggia-Rose formalism.}
The generating functional is constructed from the path integral
\beq
    Z[h] = \int\mathcal{D}R\,P[R(t)]\exp\left(\intd{t}\,h(t)\cdot R(t)\right).
\eeq
Here, the functions $h_\mu(t)$ are external sources for the degrees of freedom. 
Each path $R(t)$ is weighted by its probability. 
In order to avoid solving the stochastic differential equations explicitly, 
we employ the Martin-Siggia-Rose formalism \cite{PhysRevA.8.423}. 
Therefore, we introduce a response field $\chi_\mu(t)$ that is the Fourier conjugate to the equations of motion. 
The field $\chi_\mu(t)$ quantifies how a system responds to a delta-shaped perturbation 
of the degrees of freedom and is closely related to the susceptibility. 
In this way, we can perform the integration over the Gaussian random processes $\eta_\mu(t)$ 
and arrive at an expression for the generating functional
\beq
    \hspace{-1cm}Z[h,g] &= \int\mathcal{D}R\int\mathcal{D}\chi\int\mathrm{d}R_0\,P\left[R_0\right]\exp\bigg(-S[R,\chi]
    +\left(h|R\right)+\left(g|\chi\right)
    \bigg),\label{eq:Z[h,g]}
\eeq
which involves an average over the initial conditions $R_0$. Here 
we introduced another source field $g_\mu(t)$ associated with the response 
fields $\chi_\mu(t)$ and we defined the scalar product
\beq
    \left(h|R\right) = \intd{t}\,h(t)\cdot R(t)
\eeq 
and analogously for $(g|\chi)$. The action
\beq
    S[R,\chi] &= \intd{t}\,\left[\frac{1}{2}\chi_\mu\sigma_{\mu\gamma}\sigma_{\nu\gamma}\chi_\nu+\mathrm{i}\chi_\mu\left(\frac{\mathrm{d}R_\mu}{\mathrm{d}t\,}-F_\mu[R]\right)+G_\mathrm{R}(0)\frac{\mathrm{d}F_\mu}{\mathrm{d}R_\mu}\right]\label{eq:action-general}
\eeq
is quadratic in the response fields, which results from the random processes $\eta_\mu(t)$ being Gaussian. 
Note that the action is deterministic now, since the noise is already ``integrated out''.
The last term of the action arises from the functional Jacobian in Eq.~(\ref{eq:transition-probability}). 
A proper treatment of this determinant requires the introduction of Grassmann functions, 
as discussed in \cite{PhysRevE.91.042103} and in \ref{app1}, as they give rise to a 
Becchi-Rouet-Stora-Tyutin (BRST) symmetry and even supersymmetry. 
Their relevance in the context of classical mechanics was discussed in \cite{PhysRevD.40.3363}. 
For our discussion it suffices to say that the stochasticity of the underlying dynamics 
is encoded in the Grassmann functions, but since we have integrated them out, 
we are left with the retarded Green's function $G_\mathrm{R}(0)$ of the Grassmann fields 
evaluated at zero. In \ref{app1} we show that this is the Heaviside theta function. 
It should be noted that here also the ambiguity of stochastic prescriptions presents itself. 
For a retarded or It\^o prescription $G_\mathrm{R}(0)=0$, while for the symmetric 
Stratonovich prescription $G_\mathrm{R}(0)=1/2$, cf.\,\cite{PhysRevE.91.042103} and \cite{Barci}.

\section{Active Brownian Particle in two Dimensions}
\label{secActive}

We now specify the general first order stochastic differential equation (\ref{eq:stochastic-eom}) 
to describe an active Brownian particle (ABP) in two dimensions (2D)\footnote{In the following,
 we use vector arrows only for vectors in 2D space.}. 
Such a particle has coordinates $\vec q=(x,y)^\intercal$ and it propels itself with a constant velocity $v_0$ 
in the direction $\vec n(\vartheta)=(\cos\vartheta,\sin\vartheta)^\intercal$ given by the direction angle $\vartheta$. 
Setting the friction coefficients to one, the equations of motion are
\beq
    \derivative{t}{\vec q} &= v_0\vec n(\vartheta)-\alpha\vec q+\sqrt{2D_q}\,\vec\eta_q\,,\label{qeq}\\
    \derivative{t}{\vartheta} &= \sqrt{2D_\vartheta}\,\eta_\vartheta\,.
\eeq
The random processes $\eta_\mu(t)$ have the form of Gaussian white noise and obey
\beq
\left<\eta_\mu(t)\right>=0,\qquad\left<\eta_\mu(t)\eta_\nu(t')\right>=\delta_{\mu\nu}\delta(t-t'),
\eeq
where Greek indices label $\{x,y,\vartheta\}$. The diffusion coefficient $D_\vartheta$ (also called  $D_{R}$) 
is the inverse of the timescale $\tau_\mathrm{R}$ of the rotational diffusion and scales for 
a spherical particle like its inverse volume.
Translational diffusion $D_q$
scales for a spherical particle like its inverse radius  \cite{RevModPhys.88.045006}.

We assume the particle is placed in a rotationally symmetric harmonic trap given 
by a potential $V(q)=\alpha q^2/2$, with $\alpha>0$ the stiffness of the trap.
One can already anticipate an interesting feature of the trapped ABP:
without activity ($v_0=0$), Eq.~(\ref{qeq}) corresponds to the Ornstein-Uhlenbeck process
and in the long-time limit yields a Gaussian probability distribution around the center of 
the trap with a finite width of
$L_\mathrm{OUP}=\sqrt{D_q/\alpha}$. 
In contrast, its activity allows the ABP to move to a finite radius of order $v_0/\alpha$,
where propulsion and restoring forces balance. 
Hence, for sufficiently strong activity,
%depending on parameters, 
the probability distribution at a finite radius can be larger than in the center,
as discussed in more detail in section \ref{secResults}. 

The generating functional is given by Eq.~(\ref{eq:action-general}) 
and we can use the equations of motion to specify the action. 
Introducing the response fields $\vec\chi_q=(\chi_x,\chi_y)^\intercal$ and $\chi_\vartheta$ 
for the position and the direction angle, respectively, it takes the form
\beq
    \hspace{-1.5cm}S &= \intd{t}\,\left[D_q\chi_q^2+\mathrm{i}\vec\chi_q\cdot\left(\derivative{t}{\vec q}+\alpha\vec q-v_0\vec n(\vartheta)\right)+D_\vartheta\chi_\vartheta^2+\mathrm{i}\chi_\vartheta\derivative{t}{\vartheta}-G_\mathrm{R}(0)\alpha\right].
\eeq
It should be noted that position and direction angle are only coupled via the self-propulsion. 
In the rest of the discussion we will drop the last term, $-G_\mathrm{R}(0)\alpha$, 
as it is a constant and may be absorbed in the normalisation of the generating functional $Z[h,g]$.

In order for the path integral to be fully defined we need to specify the initial distribution of the degrees of freedom as well. Throughout this work we assume
\beq
    P\left[\vec q_0,\vartheta_0\right]=P\left(\vec q_0\right)P\left(\vartheta_0\right),\label{eq:uncorrelated-IC}
\eeq
meaning that the position of the particle and its direction of self-propulsion are not correlated at the initial time.

\subsection{The reference motion}

We have now formulated the problem as a path integral for the generating function
and would like to find solutions for the degrees of freedom $(\vec q,\vartheta)$ 
and the response fields $(\vec\chi_q,\chi_\vartheta)$. 
An analytical solution to the full dynamics 
is not always available, and then perturbative methods need to be applied. 
The strategy is defining a ``free'' or reference motion, such as ballistic motion in the case of Newtonian dynamics, 
and treat any deviation from that reference perturbatively. 
Obviously, it is advantageous for the reference motion to be as close to the actual dynamics as possible
for the  perturbative expansion to be efficient.

In the case of an ABP we choose the activity to be the deviation from the reference motion. 
The reference motion is then an Ornstein-Uhlenbeck process for the position $\vec q$ 
and a Brownian motion for the direction angle $\vartheta$. 
Thus, we separate the action into two parts
\beq
    S_0 &=\intd{t}\,\left[D_q\chi_q^2+\mathrm{i}\vec\chi_q\cdot\left(\derivative{t}{\vec q}+\alpha\vec q\right)+D_\vartheta\chi_\vartheta^2+\mathrm{i}\chi_\vartheta\derivative{t}{\vartheta}\right],\\
    S_A &= -\mathrm{i}v_0\intd{t}\,\left[\vec\chi_q\cdot\vec n(\vartheta)\right].
\eeq
In this expression $S_0$ refers to the reference motion and $S_A$ 
represents the particle's self-propulsion. We may replace 
\beq
    \vartheta(t)\rightarrow\funcderiv{h_\vartheta(t)}{},\quad\mathrm{and}\quad\vec\chi_q(t)\rightarrow\funcderiv{\vec g_q(t)}{}
\eeq
to pull the factor $\exp(-S_A)$ as an operator in front of the path integral
\beq
    Z_\mathrm{ABP}[h,g] = \exp(-{S_A})Z_0[h,g].
    \label{eq:Z_ABP}
\eeq
In the context of a quantum field theory, one would refer to $S_A$ as the ``interaction operator''. 
Here we rather call it the ``activity operator'' and reserve the term ``interaction'' for higher order 
potential interactions or pairwise interactions in systems of many active particles (to be studied in the future). 
Since the reference motion does not couple the position $\vec q$ and the direction angle $\vartheta$,
 we can define the free generating functional as a product
\beq
    Z_0[h,g] &= Z_0^\alpha\left[\vec h_q,\vec g_q\right]\cdot Z_0^\vartheta\left[h_\vartheta,g_\vartheta\right].
\eeq
Then the free generating functional for the position is 
\beq
  \hspace{-1cm}  Z_0^\alpha\left[\vec h_q,\vec g_q\right]&=\int\mathcal{D}\vec q\int\mathcal{D}\vec\chi_q\intd{^2q_0}\,P(\vec q_0)\exp\left[-S_0^\alpha+(\vec h_q| \vec q)+(\vec g_q|\vec\chi_q)\right]
\eeq
with the Ornstein-Uhlenbeck action 
\beq
    S_0^\alpha &= \intd{t}\,\left[D_q\chi_q^2+\mathrm{i}\vec\chi_q\cdot\left(\derivative{t}{\vec q}+\alpha\vec q\right)\right].
\eeq
In turn, the generating functional for the direction angle is
\beq
   \hspace{-1.5cm} Z_0^\vartheta\left[h_\vartheta,g_\vartheta\right] &=\int\mathcal{D}\vartheta\int\mathcal{D}\chi_\vartheta\intd{\vartheta_0}\,P(\vartheta_0)\exp\bigg[-S_0^\vartheta
   +\left(h_\vartheta |\vartheta\right)+\left(g_\vartheta|\chi_\vartheta\right)\bigg]
\eeq
with the action for Brownian motion
\beq
    S_0^\vartheta &= \intd{t}\,\left[D_\vartheta\chi_\vartheta^2+\mathrm{i}\chi_\vartheta\derivative{t}{\vartheta}\right].
\eeq
The free generating functional $Z_0[h,g]$ contains all information about the reference motion, i.e.\,a passive particle in two dimensions undergoing translational and rotational diffusion while being trapped in a harmonic potential 
with stiffness $\alpha>0$. 

By functional differentiation of $Z_0^\alpha$ with respect to $\vec q$ and $\vec\chi_q$, one can now derive 
the classical equations of motion for the {\it expectation values} of the position and the associated response 
field for fixed initial position, i.e.~$P[\vec q_0]=\delta(\vec q-\vec q_0)$.
They read
\begin{eqnarray}
\label{Schwingerdyson1}
    \quad\mathrm{i}\left(\derivative{t}{}+\alpha\right)\left<\vec q\right>
    &=&\vec g_q-2D_q\left<\vec\chi_q\right>,\\
\label{Schwingerdyson2}    
    -\mathrm{i}\left(\derivative{t}{}-\alpha\right)\left<\vec\chi_q\right>&=&\vec h_q.
\end{eqnarray}
These are the Schwinger-Dyson equations for an Ornstein-Uhlenbeck process. Their solutions are readily obtained as
\beq
    \left<\vec\chi_q(t)\right>&=\inte{\tau}{t_0}{\infty}\,G_q\left(\tau,t\right)h_\nu(\tau),\label{eq:chi-q(t)}\\
    \left<\vec q(t)\right> &= \mathrm{i}G_q(t,t_0)\vec q_0 + \inte{\tau}{t_0}{\infty}\,G_q(t,\tau)\vec g_q(\tau)+\inte{\tau}{t_0}{\infty}\,C_q(t,\tau)\vec h_q(\tau),\label{eq:q(t)}
\eeq
where $t_0$ refers to the initial time. For the initial/boundary conditions we use 
$\vec q(t_0)=\vec q_0$ and that the response field is decaying after 
the system was disturbed, $\vec\chi_q(t\rightarrow\infty)=0$. As usual we call the functions $G_q(t,\tau)$ and $C_q(t,\tau)$ the 
``propagators'' and these are explicity given by
\beq
    G_q(t,\tau) &= -\mathrm{i}\Theta(t-\tau)\exp\left[-\alpha(t-\tau)\right],\\
    C_q(t,\tau)&=-2D_q\inte{\bar{t}}{t_0}{\infty}\,G_q(t,\bar{t})\,G_q(\tau,\bar{t})=C_q(\tau,t).
\eeq
In particular, note that the propagator $G_q(t,\tau)$ solves the Green's function equation
\beq
    \left(\pderiv{t}{}+\alpha\right)G_q(t,\tau)=-\mathrm{i}\delta(t-\tau).\label{eq:Green-equation}
\eeq

With the solutions (\ref{eq:chi-q(t)}) and (\ref{eq:q(t)}) at hand, we can make use of the fact that 
the expectation values may be calculated by functional derivatives of the generating functional
with respect to the source fields, such as
\beq
    \left<\vec q(t)\right> = \funcderiv{\vec h_q(t)}{}\ln Z_0^\alpha\left[\vec h_q,\vec g_q\right].
\eeq
Therefore, the free generating functional for the Ornstein-Uhlenbeck process can be written as
\beq
   \hspace{-2.5cm} Z_0^\alpha\left[\vec h_q,\vec g_q\right] 
   &= \exp\left[\frac{1}{2}\hspace{-1mm}\left(\vec h_q\right|C_q\left|\vec h_q\right)+\left(\hspace{-1mm}\left.\left.\vec h_q\right|G_q\right|\vec g_q\right)\right]
    \intd{^2q_0}\,P\left(\vec q_0\right)\exp\left[\mathrm{i}\left(\hspace{-1mm}\left.\left.\vec h_q\right|G_q\right|\vec q_0\right)\right]\hspace{-1mm}.
\eeq
In order to unclutter notations we have introduced another scalar product
\beq
    \left(\vec h_q\right|C_q\left|\vec h_q\right) &= \inte{\tau}{t_0}{\infty}\,\inte{\tau'}{t_0}{\infty}\,\vec h_q(\tau)\cdot C_q(\tau,\tau')\vec h_q(\tau'),
\eeq
and analogously for $\left(\hspace{-1mm}\left.\left.\vec h_q\right|G_q\right|\vec g_q\right)$, as well as
$\left(\hspace{-1mm}\left.\left.\vec h_q\right|G_q\right|\vec q_0\right) = \inte{\tau}{t_0}{\infty}\,\vec h_q(\tau)\cdot G_q(\tau,t_0)\,\vec q_0$.
We can do the same calculation for the direction angle and the corresponding response field.
 The structure of the solution is the same, namely
\beq
  \hspace{-2cm}  Z_0^\vartheta[h_\vartheta,g_\vartheta] &= \exp\left[\frac{1}{2}\left(h_\vartheta|C_\vartheta|h_\vartheta\right)+\left(h_\vartheta|G_\vartheta|g_\vartheta\right)\right]
    \intd{\vartheta_0}\,P(\vartheta_0)\exp\left[\mathrm{i}\left(h_\vartheta|G_\vartheta|\vartheta_0\right)\right].
\eeq
Since there is no  potential for the directional degree of freedom, the propagators take on the simpler form
\beq
    G_\vartheta(t,\tau) &=-\mathrm{i}\Theta(t-\tau)\,,\\
    C_\vartheta(t,\tau) &=2D_\vartheta\inte{\bar t}{t_0}{t}\inte{\tau'}{\bar t}{\infty}\,\delta(\tau-\tau')\,.
\eeq
This result for the Wiener process was already given in \cite{PhysRevE.81.061102}. 
In particular, the diffusion propagator $C_\vartheta(t,\tau)$ satisfies
\beq
    C_\vartheta(t,t) = 2D_\vartheta (t-t_0)
    \label{eq:C_theta(t,t)}
\eeq
for equal times. With these expressions it is straightforward to calculate 
all moments of the probability distribution and the distribution itself 
for a passive Brownian particle in a harmonic potential, as shown in section \ref{secResults}
as the limiting case of vanishing activity/self-propulsion.

\subsection{Activity / Self-Propulsion}
\label{ops_define}
The deviations from the reference motion are caused by the particle's self-propulsion,
hence for an active Brownian particle with constant speed $v_0\neq 0$ we need the 
full generating functional as given in Eq.~(\ref{eq:Z_ABP}). 
For convenience, we redefine the activity operator representing these deviations from
the reference motion as
\beq
    {S_A} &= -\mathrm{i}v_0{\hat A}\,,\,\,\,{\rm with}\,\,\,{\hat A}\left[\funcderiv{h_\vartheta}{},\funcderiv{\vec g_q}{}\right]=\frac{1}{2}\left\{(a^+|\phi^+)+(a^-|\phi^-) \right\}
\eeq
a new operator.
In this expression, 
Euler's formula was used to rewrite the trigonometric functions and we introduced the set of operators
\beq
     a^{\sigma}(t)= \frac{\delta}{\delta g_x(t)} + \mathrm{i}\sigma \frac{\delta}{\delta g_y(t)} \quad \mathrm{and} \quad \phi^{\sigma}(t)=\ex^{- \mathrm{i}\sigma\frac{\delta}{\delta h_{\vartheta}(t)}}
\eeq
with $\sigma\in\{+,-\}$.  Applying the operators 
$\phi^\sigma(t)$ and $a^\sigma(t)$ 
to the free generating functional $Z_0[h,g]$ yields
\beq
    \phi^{\sigma}(t)Z_0[h,g] &=Z^{\alpha}_0\left[\vec{h}_q,\vec{g}_q\right]\cdot Z^{\vartheta}_0[h_{\vartheta}-\mathrm{i}\sigma\delta_t, g_{\vartheta}],\label{eq:phi-operator}\\
    a^{\sigma}(t)Z_0[h,g] &= \left\{(hG_q)_x(t)+\mathrm{i}\sigma(hG_q)_y(t)\right\}\cdot Z_0[h,g],\label{eq:a-operator}
\eeq
with the notation $\delta_t=\delta(t-t')$ and where the integral
\beq
    \left(hG_q\right)_i(t)=\intd{t}'\,h_i(t')G_q(t',t)
\eeq
with $i\in\{x,y\}$ was introduced.

As already stated, for many practical calculations an analytical solution of the full problem is not available. 
Nevertheless, by means of a Taylor expansion, one can determine the effect of the particle's self-propulsion 
in a perturbative series:
expectation values $\mathcal{O}(\vec q,\vartheta)$ can be calculated order by order,
\beq
    \mean{\mathcal{O}(\vec{q},\vartheta)} = \mean{\mathcal{O}(\vec{q},\vartheta)}^{(0)} + \sum_{N=1}^{\infty} \Delta^{(N)} \mean{\mathcal{O}(\vec{q},\vartheta)},
\eeq
where $\mean{\mathcal{O}}^{(0)}$ is the average with respect to the free generating functional $Z_0[h,g]$ and the corrections of order $N$ are given by
\beq \label{series_in_v}
    \Delta^{(N)} \mean{\mathcal{O}(\vec{q},\vartheta)} = \left.\frac{(\mathrm{i}v_0)^N}{N!}\mathcal{O}\left(\frac{\delta}{\delta \vec{h}_q},\frac{\delta}{\delta h_{\vartheta}}\right)\left[{\hat A}^NZ_0[h,g]\right]\right|_{h,g=0}.\label{eq:perturbative-correction}
\eeq
Here we are especially interested in the average position (related to $N=1$) and the mean square displacement 
(MSD, related to $N=2$)
of a trapped ABP, because those are important from an experimental point of view and 
can be compared to previously obtained results \cite{PhysRevLett.129.158001}.

\section{Results}
\label{secResults}

In the previous section we have formulated a generating functional $Z_\mathrm{ABP}[h,g]$ 
for an active Brownian particle in an isotropic harmonic trap in two dimensions, 
based on its microscopic degrees of freedom, namely the position $\vec q(t)$ and 
orientation angle $\vartheta(t)$. We have separated the self-propulsion as a deviation 
from the motion of a passive Ornstein-Uhlenbeck particle (which includes the trap)
that additionally undergoes rotational Brownian motion. 
Based on this formalism,  in this section we derive analytic expressions for the mean position, 
the mean-square displacement (MSD) and the probability distribution of the particle's position. 
All results are given for arbitrary distributions of initial conditions satisfying
Eq.~(\ref{eq:uncorrelated-IC}), i.e.~without correlations between initial 
positions and direction angles, but we note that the approach could also deal with such correlations.
For simplicity we assume that the initial time satisfies $t_0=0$.

\subsection{Average Position}

The average position at time $t$ is given by a single functional derivative 
with respect to the source $\vec h_q(t)$:
\beq
    \mean{\vec{q}(t)}= \left.\funcderiv{\vec{h}_q(t)}{} Z_\mathrm{ABP}[h,g]\right|_{h, g=0}\,,
\eeq
which we calculate perturbatively in orders of the self-propulsion velocity $v_0$. 
At zeroth order, %$\mathcal{O}(1)$, 
we recover our previous result, Eq.~(\ref{eq:q(t)}), with vanishing sources
\beq
\hspace{-1cm}\mean{\vec{q}(t)}^{(0)}=\left.\funcderiv{\vec{h}_q(t)}{} Z_{0}[h,g]\right|_{h, g=0}= \intd{^2q_0}\,P\left(\vec q_0\right)\mathrm{i}G_q(t,0)\vec q_0
=  \mean{\vec{q}_0} \ex^{-\alpha t}\,,
\eeq
which is simply the relaxation of the passive Brownian particle in the harmonic trap,
with the characteristic time $1/\alpha$ (note that friction/mobility was put to one).

In order to calculate the first order correction $ \Delta^{(1)}\mean{\vec{q}(t)}$
in $\mathcal{O}(v_0)$ we apply the  activity operator $S_A=-\mathrm{i}v_0{\hat A}$ once, 
before taking the functional derivative 
with respect to $\vec{h}_q(t)$. To do so, we first evaluate the $x$-direction
\beq
  \hspace{-1.5cm}  &\left.\frac{\delta}{\delta h_x(t)}\left[(a^\sigma|\phi^\sigma)Z_0[h,g]\right]\right|_{h,g=0} \nonumber\\
  \hspace{-1.5cm}   &= \left.\frac{\delta}{\delta h_x(t)} \left[\int_0^\infty\mathrm{d}t' \left( (hG_q)_x(t')+(hG_q)_y(t')\right)Z_0^\alpha[\vec{h},\vec{g}]Z_0^\vartheta[h_\vartheta-\mathrm{i}\sigma\delta_t,g_\vartheta]\right]\right|_{h,g=0}\nonumber\\
  \hspace{-1.5cm}   &=\left.\int_0^\infty\mathrm{d}t'\left(G_q(t,t')+\mathcal{O}(\vec{h})\right)Z_0^\alpha[\vec{h},\vec{g}]Z_0^\vartheta[h_\vartheta-\mathrm{i}\sigma\delta_t,g_\vartheta]\right|_{h,g=0}\nonumber\\
  \hspace{-1.5cm}    &=\int_0^\infty\mathrm{d}t' G_q(t,t') Z_0^\vartheta[-\mathrm{i}\sigma\delta_t,0],
\eeq
where again $\sigma\in\{+,-\}$. Following the same steps leads to a similar expression for the $y$-direction
\beq
    \left.\frac{\delta}{\delta h_y(t)}\right|_{h=g=0}\left[(a^\sigma|\phi^\sigma)Z_0[h,g]\right] = \mathrm{i}\sigma \int_0^\infty\mathrm{d}t' G_q(t,t') Z_0^\vartheta[-\mathrm{i}\sigma\delta_t,0].
\eeq
With the diffusion propagator $C_\vartheta(t',t')=2D_\vartheta t'$ one easily calculates %by plugging in
%\beq
$Z_0^\vartheta[-\mathrm{i}\sigma\delta_t,0]=\ex^{-\mathrm{i}\sigma\vartheta_0-D_\vartheta t'}$
%\eeq
and with the integral
$\int_0^\infty\mathrm{d}t' G_q(t,t')\ex^{-\mathrm{i}D_\vartheta t'} 
%= \mathrm{i}\int_0^\infty \mathrm{d}t' \theta(t-t')\ex^{-D_\vartheta t'+\alpha(t'-t)} 
= \frac{\mathrm{i}}{\alpha-D_\vartheta}(\ex^{-D_\vartheta t}-\ex^{-\alpha t}),
$
using Eq.~(\ref{series_in_v}) we get the correction of the expectation value at linear order $\mathcal{O}(v_0)$ as
\beq
  \hspace{-2cm}  \Delta^{(1)}\mean{q_x(t)} &= \frac{\mathrm{i}v_0}{2}\int_0^\infty\mathrm{d}t'G_q(t,t')
  \bigg[Z_0^\vartheta[-\mathrm{i}\delta_t,0]+Z_0^\vartheta[+\mathrm{i}\delta_t,0]\bigg] \nonumber\\
  &= \frac{v_0}{\alpha - D_{\vartheta}}\left(\ex^{-D_\vartheta t}-\ex^{-\alpha t}\right)\mean{\cos(\vartheta_0)}, \\
 \hspace{-2cm}   \Delta^{(1)}\mean{q_{ y}(t)} &= \frac{v_0}{2}\int_0^\infty\mathrm{d}t'G_q(t,t') 
  \bigg[Z_0^\vartheta[-\mathrm{i}\delta_t,0]-Z_0^\vartheta[+\mathrm{i}\delta_t,0]\bigg] \nonumber\\
  &= \frac{v_0}{\alpha - D_{\vartheta}}\left(\ex^{-D_\vartheta t}-\ex^{-\alpha t}\right)\mean{\sin(\vartheta_0)}.
\eeq
Here, $\mean{\dots}$ on the r.h.s.~denotes the angular average with respect to the initial distribution. 
One can see that the correction at linear order does not depend on the initial position, but on the initial orientation. 
Importantly, higher orders $\mathcal{O}(v_0^N)$ with $N\geq 2$ of the perturbative series vanish exactly. This can be seen 
from the form of the perturbative corrections, Eq.~(\ref{eq:perturbative-correction}): The $N$-th order correction 
is proportional to the source $\vec h_q^{N}$. Thus, setting the source fields $h$ and $g$ to zero after having applied 
the functional derivative with respect to $\vec h_q$, leads to vanishing corrections at that order.

Therefore, although performing a perturbation series, we derived an {\it exact expression} for the mean position 
of an ABP in an isotropic harmonic potential, 
\beq\label{eq:meanq}
    \mean{\vec{q}(t)}=\ex^{-\alpha t}\mean{\vec{q}_0}+\frac{v_0\ex^{-\alpha t}}{D_\vartheta - \alpha}\left(1-\ex^{-(D_{\vartheta}-\alpha)t}\right)\mean{\vec{n}_0}.
\eeq
This result was also given in \cite{PhysRevLett.129.158001}.
At short times, the particle starting at $\vec{q}_0$ will drift along $\vec{n}_0$,  
but in the end the average position always decays to zero.
Our result is also compatible with the free ABP (without the harmonic potential), 
as given e.g.~in \cite{RevModPhys.88.045006}
and obtained from Eq.~(\ref{eq:meanq}) in the limit 
$\alpha\rightarrow0$ as
%\beq
$    \mean{\vec{q}}_{\alpha=0} = \mean{\vec{q}_0} + \frac{v_0}{D_{\vartheta}}(1-\ex^{-D_{\vartheta}t})
\mean{\vec{n}_0}$.
%\eeq
There, the particle propels a distance of order $v_0/D_{\vartheta}$
before the orientation decorrelates on the time scale $1/D_{\vartheta}$.
In a sense, in the case with potential, the potential always ``wins'', and the relaxation is more complex
involving two timescales, $1/\alpha$ and $1/D_{\vartheta}$.

\subsection{The Mean Square Displacement}

The mean square displacement 
\beq
\mathrm{MSD}(t) =\mean{\vec{q}^{\,2}(t)}-2\mean{\vec{q}(t) \cdot \vec q_0 }+\mean{\vec{q}_0^{\,2}}
\eeq
can also be calculated 
exactly from the generating functional $Z_\mathrm{ABP}[h,g]$. 
The second term is readily obtained from our previous result, Eq.~(\ref{eq:meanq}). 
Thus, we only have to calculate $\mean{\vec{q}^{\,2}(t)} = \mean{q_x^2(t)}+\mean{q_y^2(t)}$ 
by taking the second functional derivative with respect to the source field $\vec h_q(t)$. 
At zeroth order % $\mathcal{O}(1)$, 
we get
\beq
    \hspace{-1cm}\mean{q_i(t)q_j(t)}^{(0)} &= \frac{\delta^2Z_0[h,g]}{\delta h_i(t) \delta h_j(t)}% \\ &
    = \frac{D_q}{\alpha}\left(1-\ex^{-2\alpha t} \right)\delta_{ij} + \mean{\left(\ex^{-\alpha t}q_{0,i}\right)\left(\ex^{-\alpha t}q_{0,j}\right)}
\eeq
with $i,j\in\{x,y\}.$ Contraction over the indices leads to 
\beq
    \mean{\vec{q}^{\,2}(t)}^{(0)}=\frac{2D_q}{\alpha}(1-\ex^{-2\alpha t}) + \ex^{-2\alpha t}\mean{\vec{q}_0^{\,2}}.
\eeq
As expected, we recover the result for an Ornstein-Uhlenbeck process. 
For long times, the initial condition does not matter and the particle
diffuses on the characteristic length scale $L_\mathrm{OUP}=\sqrt{D_q/\alpha}$ 
in the $x$- and $y$-directions.

To determine the first order correction at $\mathcal{O}(v_0)$ to $\mean{\vec{q}^2(t)}$, 
we calculate the corrections for $\mean{q_x^2(t)}$ and $\mean{q_y^2(t)}$ separately. 
For the $x$-direction we need to evaluate
\beq
    \Delta^{(1)}\mean{q_x^2(t)}%&=\frac{\delta^2 (-\mathrm{i}v_0 {\color{blue}{\hat A}})Z_0[h,g]}{\delta h_x(t)\delta h_x(t)}\\
    &= \frac{\mathrm{i}v_0}{2} \frac{\delta^2}{\delta h_x(t) \delta h_x(t)} \left[\left(a^+|\phi^+\right)+\left(a^-|\phi^-\right)\right] Z_0[h,g] \Bigg|_{h,g=0}.
\eeq
Using the expressions (\ref{eq:phi-operator}) and (\ref{eq:a-operator}) we find
\beq
\hspace{-.5cm}\Delta^{(1)}\mean{q_x^2(t)} 
&= \mathrm{i} v_0 \int_0^t \mathrm{d}t' \, G_q(t,t') \mean{q_x(t)}^{(0)} \left( Z_0^\vartheta[-\mathrm{i}\delta_{t'},0]+Z_0^\vartheta[+\mathrm{i}\delta_{t'},0] \right),
\eeq
where we explicitly used that initial positions and orientations are independently distributed. 
In the same way as in the previous section, we find 
\beq
    \Delta^{(1)}\mean{q_x^2(t)} &= 2\mathrm{i}v_0 \cos(\vartheta_0) \mean{q_x(t)}^{(0)} \int_0^t \mathrm{d}t' G_q(t,t') \ex^{-D_{\vartheta}t'}\nonumber\\
    &= \frac{2v_0}{\alpha -D_{\vartheta}} \left(\ex^{-D_{\vartheta}t}-\ex^{-\alpha t}\right) \mean{\cos(\vartheta_0)}\mean{q_x(t)}^{(0)}.
\eeq
An almost identical calculation leads to a similar expression for the $y$-direction
\beq
    \Delta^{(1)}\mean{q^2_y(t)} = \frac{2v_0}{\alpha -D_{\vartheta}}\left(\ex^{-D_{\vartheta}t}-\ex^{-\alpha t}\right) \mean{\sin(\vartheta_0)} \mean{q_y(t)}^{(0)}.
\eeq
To find the linear correction to $\mean{\vec q^{\,2}(t)}$ we sum the contributions from the $x$- and $y$-directions and use $\mean{\vec q(t)}^{(0)}=\ex^{-\alpha t}\mean{\vec{q}_0}$, which leads to 
\beq
    \Delta^{(1)}\mean{\vec q^{\,2}(t)} =\frac{2v_0}{\alpha - D_{\vartheta}} \left(\ex^{-(D_{\vartheta} + \alpha)t}-\ex^{-2\alpha t}\right) \mean{\vec{n}(\vartheta_0)}\cdot\mean{\vec{q}_0}.
\eeq
The linear correction is therefore explicitly dependent on the initial conditions. 
In particular, if positions and orientation were not independent initially, this correction would be proportional
to the correlation function $\mean{\vec n(\vartheta_0)\cdot\vec q_0}$ instead of $ \mean{\vec{n}(\vartheta_0)}\cdot\mean{\vec{q}_0}$.

Next we need to evaluate the quadratic order $\mathcal{O}(v_0^2)$. Again, we start by writing down the corrections for the $x$- and $y$-components separately and get an expression
\beq
    \hspace{-2.5cm}\Delta^{(2)}\mean{q_i^2(t)} &= \left(\frac{\delta}{\delta h_i(t)}\right)^2\left.\left[-\frac{v_0^2}{2}{\hat A}^2 Z_0[h,g]\right]\right|_{h,g=0} \nonumber\\
    \hspace{-1cm}&= -\frac{v_0^2}{8} \left(\frac{\delta}{\delta h_i(t)}\right)^2 \bigg[\big(a^+|\phi^+\big)^2 + 2\big(a^+|\phi^+\big) \big(a^-|\phi^-\big) + \big(a^-|\phi^-\big)^2 \bigg] Z_0[h,g] \Bigg|_{h,g=0}
\eeq
with $i\in\{x,y\}$. For notational convenience we define 
\beq
    (\sigma\varepsilon) %&:= \left(\frac{\delta}{\delta h_x(t)}\right)^2 \left[\left(a^{\sigma}|\phi^{\sigma}\right)\left(a^{\varepsilon}|\phi^{\varepsilon}\right)Z_0[h,g]\right]\\
    &= \int_{t'}\int_{t''} G_q(t,t')G_q(t,t'')Z_0^{\vartheta}[-\mathrm{i}\sigma\delta_{t'}-\mathrm{i}\varepsilon\delta_{t''},0]
\eeq
where $\sigma,\varepsilon\in\{+,-\}$. 
Then we can write the quadratic corrections in short-hand forms
\beq
    \Delta^{(2)}\mean{q_x^2(t)} &= -\frac{v_0^2}{8} \left[(++)+4(+-)+(--)\right], \\
     \Delta^{(2)}\mean{q_y^2(t)} &= +\frac{v_0^2}{8} \left[(++)-4(+-)+(--)\right].
\eeq
Adding both corrections yields for the total correction at second order 
$\Delta^{(2)}\mean{\vec{q}^2(t)} = -\frac{v_0^2}{2}(+-)$, such that only the integral $(+-)$ needs to be evaluated. 
The calculation is straightforward, but a bit tedious and detailed in \ref{app3}. The full result is
\beq
\Delta^{(2)}\mean{\vec{q}^{\,2}(t)}=\frac{2v_0^2}{\alpha+D_{\vartheta}} \left\{\frac{1-\ex^{-2\alpha t}}{2\alpha}-\frac{\ex^{-(\alpha+D_{\vartheta})t}-\ex^{-2\alpha t}}{\alpha-D_{\vartheta}} \right\}.
\eeq
Again, the perturbative series terminates at this second order and all corrections $\Delta^{(N)}\mean{\vec q^2(t)}$ 
with $N\geq 3$ are exactly zero. The argument is similar to before 
and can be understood from the solution (\ref{eq:q(t)}) of the Schwinger-Dyson equation: 
this solution is linear in the source field $\vec g_q(t)$, such that the position will only experience perturbative 
corrections up to linear order. For a general function of the particle position, that can be written as a power-law 
of order $n$, the perturbative series will hence terminate at order $n$ and corrections with $N>n$
are identically zero.

Collecting all terms we find the exact analytic expression for the MSD of an ABP in a harmonic potential
\beq
    \hspace{-2.5cm}\mathrm{MSD(t)} &= \left[1-\ex^{-\alpha t}\right]^2\mean{\vec q_0^2}+\frac{2v_0}{D_\vartheta-\alpha}\left[1-\ex^{-(D_\vartheta-\alpha)t}\right]\left[\ex^{-2\alpha t}-\ex^{-\alpha t}\right]\mean{\vec q_0}\cdot\mean{\vec n_0}\nonumber\\
    \hspace{-1.75cm}&\ +\left[\frac{2D_q}{\alpha}+\frac{D_\vartheta}{\alpha}\frac{v_0^2}{D_\vartheta^2-\alpha^2}\right]\left[1-\ex^{-2\alpha t}\right]+\frac{v_0^2}{D_\vartheta^2-\alpha^2}\left(2\ex^{-(D_\vartheta+\alpha)t}-\ex^{-2\alpha t}-1\right).\label{fullMSD}
\eeq
This is again comparable with the results from \cite{PhysRevLett.129.158001}, 
but here we have not yet specified the initial distribution. 
For trivial initial conditions (the subscript '0' indicating that the particle starts at the origin
in an arbitrary direction), the MSD is just given by the second line of Eq.~(\ref{fullMSD}), i.e. 
\beq\label{MSDtrivial}
    \hspace{-2.5cm}\mathrm{MSD}_0\left(t\right)=\left[\frac{2D_q}{\alpha}+\frac{D_\vartheta}{\alpha}\frac{v_0^2}{D_\vartheta^2-\alpha^2}\right]\left[1-\mathrm{e}^{-2\alpha t}\right]+\frac{v_0^2}{D_\vartheta^2-\alpha^2}\left[2\mathrm{e}^{-(D_\vartheta+\alpha)t}-\mathrm{e}^{-2\alpha t}-1\right].
\eeq
This result is again compatible with the free ABP, as given e.g.~in \cite{RevModPhys.88.045006},
in the limit $\alpha\rightarrow 0$.
 
 \begin{figure}[t!] 
    \centering
    \includegraphics[width=.5\linewidth]{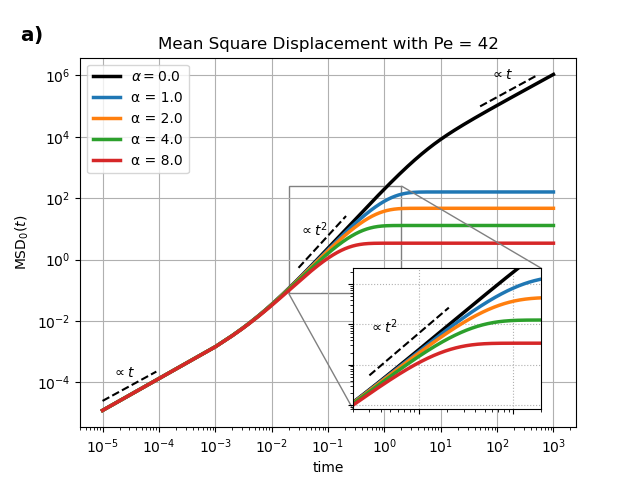}\includegraphics[width=.5\linewidth]{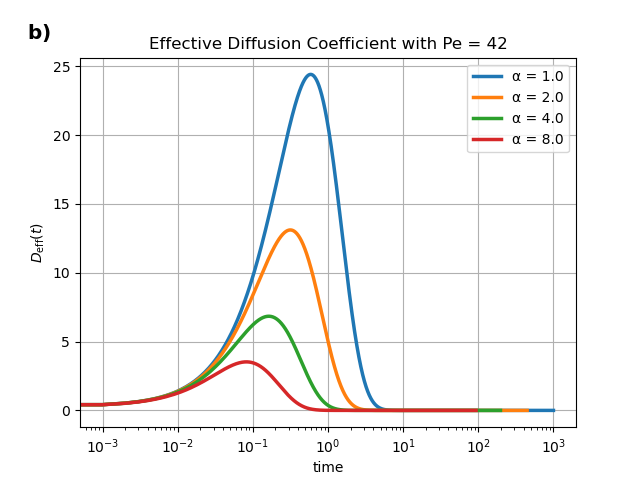}
    \caption{
    a) The mean square displacement MSD$_0(t)$ 
    and b) the effective diffusion coefficient $D_\mathrm{eff}(t)$ 
    for  an ABP in an isotropic harmonic potential with initial conditions 
    $\vec{q}_0=0$ and arbitrary direction angle $\vartheta_0\in[0,2\pi)$. 
    The stiffness of the potential $\alpha$ is varied while keeping the 
    P\'eclet number $\mathrm{Pe}=v_0/\sqrt{D_q\cdot D_\vartheta}$ constant. 
    In a) we also show the MSD for a free ABP (black curve). 
    Parameters are chosen for an experimentally realistic situation for colloidal swimmers,
    with $v_0 = 15~\mu\mathrm{m/s}$ and 
    diffusion coefficients $D_q=0.27~\mathrm{\mu m^2/s^2}$ 
    and $D_\vartheta=0.42~\mathrm{rad^2/s}$ (corresponding to a sphere of 
    radius $R=0.74~\mu\mathrm{m}$ in water at room temperature obeying 
    the Stokes-Einstein relation). }
    \label{fig:MSD and D_eff ABP}
\end{figure}

Fig.~\ref{fig:MSD and D_eff ABP} shows the result for the  trivial initial condition,
Eq.~(\ref{MSDtrivial}), together with the effective diffusion coefficient
\beq
    D_\mathrm{eff}(t)=\frac{1}{4} \frac{\mathrm{d}}{\mathrm{d} t}\mathrm{MSD} (t).
\eeq
In the figure we have chosen parameters for experimentally realistic conditions
for active colloids \cite{RevModPhys.88.045006}, 
i.e.~a spherical particle in water at room temperature obeying Stokes-Einstein relation.
In the active matter literature, the dynamics is typically characterised by the so-called 
P\'eclet number, which is given by 
\begin{equation}
\mathrm{Pe}=v_0/\sqrt{D_q\cdot D_\vartheta}
\end{equation}
and for the chosen parameters amounts to $\mathrm{Pe}\simeq 42$. 
In Fig.~\ref{fig:MSD and D_eff ABP}a) one can see the 
free ABP (black curve) moving diffusively (MSD$_0\propto t$) 
at short and long times. At intermediate timescales, the motion is ballistic 
with MSD$_0\propto t^2$. 
As soon as the ABP is trapped in a harmonic potential of increasing stiffness (colored curves), 
the motion is limited by the ratio of self-propulsion speed $v_0$ 
and the stiffness $\alpha$ of the trap: 
the particle moves diffusively for short time scales and stays at a constant 
mean square distance for long times 
with, depending on parameters, a ballistic transition regime in between. 
This behavior is also visible by the effective diffusion coefficient in 
Fig.~\ref{fig:MSD and D_eff ABP}b), which grows at short time scales, 
reaches a maximum at intermediate times and then goes to zero
due to the trapping.

\subsection{Probability Distribution}

The time-dependent probability distribution can also be calculated order by order perturbatively. 
Let us start with the definition of the probability distribution and how it can be calculated 
for a system  with degrees of freedom $R_\mu$. 
The current state of the system is the trajectory $R_\mu(t)$, such that the density at an arbitrary point 
$x_\mu$  in configuration space is simply $\rho(t,x)=\delta(x-R(t))$.  
The probability distribution at the point $x$ at time $t$ is then the average of this density over all possible paths, 
\beq
    \hspace{-2.3cm}p(t,x) = \mean{\rho(t,x)} 
    = \frac{1}{Z[0,0]} \int\hspace{-1mm}\mathcal{D}R\hspace{-1mm}\int\hspace{-1mm}\mathcal{D}\chi\hspace{-1mm}\intd{R_0}P[R_0] \delta(x-R(t))\ex^{-S[R,\chi] + (h|R) + (\chi|g)}. 
\eeq
Note that $Z[0,0]$, which we wrote here for consistency, is actually 1 in our case.
Now we make use of the fact that this expression can be obtained by 
applying the operator $\delta\left(x-\frac{\delta}{\delta h(t)}\right)$ 
to the generating functional. 
In Fourier space, this density operator becomes
\beq
    \rho(t,k):=\mathcal{FT}\left[\delta\left(x-\frac{\delta}{\delta h(t)}\right)\right] =  \ \ex^{-\mathrm{i}k_{\mu}\frac{\delta}{\delta {h}_{\mu}(t)}}.
\eeq
It hence acts like a ``translation'' or ``shift'' of the source $h(t)$ when applied to the generating functional,
leading to the expression 
\beq
    \hat p (t,k)=\mean{\rho (t,k)} = \left. \ex^{-\mathrm{i}k_{\mu}\frac{\delta}{\delta h_{\mu}(t)}} Z[h,g]\right|_{h,g=0}=: Z[-\mathrm{i}\delta_tk,0],
\eeq
where we introduced $\delta_t=\delta_t(t')=\delta(t-t')$ for ease of notation.

Let us apply this now to the generating functional for the active Brownian particle. 
We are interested in the distribution of positions only, since the direction angle $\vartheta$ diffuses freely.
Therefore, we have to shift exclusively the spatial components of $h_{\mu}$ 
\beq
    \hat{p}(t,\vec{k}) = \left.Z_\mathrm{ABP}\left[\vec{h}_q - \mathrm{i}\vec{k}\delta_t,h_{\vartheta},g\right]\right|_{h,g=0} =: Z_\mathrm{ABP}\left[-\mathrm{i}\vec{k}\delta_t,0\right],
\eeq
where we shortened notation for better readability. Using $Z_\mathrm{ABP}[h,g]=\ex^{\mathrm{i}v_0 {\hat A}} Z_0[h,g]$, 
we can now calculate $\hat{p}(t,\vec{k})$ perturbatively in orders of $v_0$. 
At zeroth order we get
\beq 
    \hat p^{(0)}(t,\vec{k}) = Z_0[-\mathrm{i}\vec{k}\delta_t, 0] 
    = \exp\left(G_q(t,0)\vec{k}\cdot\vec{q}_0 - \frac{1}{2}C_q(t,t)\vec k^2\right).
\eeq
With $G_q(t,0)=-\mathrm{i}\ex^{-\alpha t}$ and $C_q(t,t) = \frac{D_q}{\alpha}(1-\ex^{-2\alpha t})$, 
we get 
\beq\label{pk0}
    \hat p^{(0)}(t,\vec{k})=\exp\left(-\mathrm{i}\ex^{-\alpha t}\vec{k}\cdot \vec{q}_0-\frac{D_q\vec{k}^2}{2\alpha}(1-\ex^{-2\alpha t})\right),
\eeq
and inverse Fourier transformation yields the probability distribution in real space
\beq
    p^{(0)}(t,\vec{x}) =\frac{\alpha}{2\pi D_q(1-\ex^{-2\alpha t})}\exp\left(-\frac{\alpha (\vec{x}-\vec{q}_0\ex^{-\alpha t})^2}{2D_q(1-\ex^{-2\alpha t})}\right).\label{eq:prob-distr-0}
\eeq
This is exactly the result for the Ornstein-Uhlenbeck process that one
can also obtain, e.g., using the Fokker-Planck approach \cite{Risken}.

The perturbative correction at linear order is calculated by application of the activity operator. Thus, we evaluate
\beq
 \hspace{-1.5cm}   \Delta^{(1)}\hat{p}(t,\vec{k})&=(\mathrm{i}v_0 {\hat A})Z_0\left[-\mathrm{i}\vec{k}\delta_t,0\right]=\frac{\mathrm{i}v_0}{2} \bigg((a^+|\phi^+)
    +(a^-|\phi^-)\bigg)Z_0\left[-\mathrm{i}\vec{k}\delta_t,0\right].
\eeq
with $(a^{\sigma}|\phi^{\sigma})$ given in section \ref{ops_define}. 
Both contributions contain integrals of the form
\beq
 &\hspace{-2cm}   (a^{\sigma}|\phi^\sigma)Z_0\left[-\mathrm{i}\vec{k}\delta_t,0\right] \nonumber\\&=  -\mathrm{i}\int\mathrm{d}t' \left\{k_xG_q(t,t')+i\sigma k_yG_q(t,t')\right\}Z_0^\vartheta[-\mathrm{i}\sigma
 \delta_{t'},0]
 %\\   & \ \ \ 
 \cdot Z^{\alpha}_0\left[-\mathrm{i}\vec{k}\delta_t,0\right],
\eeq
which can be evaluated using $G_q(t,t')= -\mathrm{i}\Theta(t-t')\exp\left[-\alpha(t-t')\right]$ to
\beq
   (a^{\sigma}|\phi^\sigma)Z_0[-\mathrm{i}\vec{k}\delta_t,0] &= -\ex^{-\mathrm{i}\sigma\vartheta_0 - \alpha t}\big(k_x+i\sigma k_y\big)\int_0^t \mathrm{d} t' \  \ex^{(\alpha-D_{\vartheta})t'}\nonumber\\
   &= -\ex^{-\mathrm{i}\sigma \vartheta_0}(k_x+i\sigma k_y)\Bigg(\frac{\ex^{-D_{\vartheta}t}-\ex^{-\alpha t}}{\alpha-D_\vartheta}\Bigg).
\eeq
Summing both contributions for $\sigma \in \{+,-\}$, the correction to  the probability density 
in first order $\mathcal{O}(v_0)$ reads
\beq\label{pk1}
\Delta^{(1)}\hat{p}(t,\vec{k}) = -\mathrm{i}v_0  \ \vec{n}(\vartheta_0)\cdot\vec{k}\Bigg( \frac{\ex^{-\alpha t}-\ex^{-D_{\vartheta}t}}{\alpha-D_{\vartheta}}\Bigg) \hat{p}^{(0)}(t,\vec{k}).
\eeq
It should be noted that this contribution is purely anisotropic and describes in lowest order how the 
particle travels due to its initial orientation. 
It vanishes in the limit $t\rightarrow\infty$, because both
the potential and the diffusion are isotropic and hence the long-time distribution  has to be so as well.

For the second order $\mathcal{O}(v_0^2)$ we have to calculate
\beq
    \hspace{-2cm}\Delta^{(2)}\hat{p}(t,\vec{k}) &=\frac{(\mathrm{i}v_0 {\hat A})^2}{2}Z_0\left[-\mathrm{i}\vec{k}\delta_t,0\right]\nonumber\\
    &=-\frac{v_0^2}{8}\bigg(\big(a^+|\phi^+\big)^2+2\big(a^+|\phi^+\big)\big(a^-|\phi^-\big)+\big(a^-|\phi^-\big)\Bigg) Z_0\left[-\mathrm{i}\vec{k}\delta_t,0\right].
\eeq
Using the definitions of $a^{\sigma}$ and $\phi^{\sigma}$ yields
\beq
\hspace{-1cm}\big(a^{\sigma}\big|\phi^{\sigma}\big)
\big(a^{\varepsilon}\big|\phi^{\varepsilon}\big)Z_0\left[-\delta_t,h_\vartheta,0\right]
= -(\sigma \varepsilon)\bigg(k_x^2 - \mathrm{i}(\sigma+\varepsilon)k_xk_y - \sigma\eta k_y^2\bigg)\,
\hat{p}^{(0)}(t,\vec{k}),
\eeq 
where $(\sigma\varepsilon)$ with $\sigma,\varepsilon\in\{+,-\}$ again abbreviating 
the expressions calculated in \ref{app3}. Adding them up, we can see that the 
correction in second order splits into an isotropic and an anisotropic part
\beq
    \hspace{-2cm}\Delta^{(2)}\hat{p}(t,\vec{k}) &= \frac{v_0^2}{4}(+-)\vec{k}^2\hat{p}^{(0)}(t,\vec k) \nonumber\\  
    &+ \frac{v_0^2}{8}\bigg(\big(k_x^2-k_y^2\big)\big[(++)+(--)\big]-2\mathrm{i}k_xk_y\big[(++)-(--)\big]\bigg)\hat{p}^{(0)}(t,\vec k).
\eeq
Inserting $(\sigma \varepsilon)$, cf.~appendix B, 
we get explicitly
\begin{eqnarray} \label{pk2}
 \hspace{-1cm}   \Delta^{(2)}\hat{p}(t,\vec{k})&=-\frac{v_0^2}{2(\alpha+D_{\vartheta})}\Bigg(\frac{1-\ex^{-2\alpha t}}{2\alpha}-\frac{\ex^{-(\alpha+D_{\vartheta})t}-\ex^{-2\alpha t}}{\alpha-D_{\vartheta}} \Bigg){\vec{k}^2}\,\hat{p}^{(0)}(t,\vec{k})\nonumber\\
    &-\frac{v_0^2}{2(\alpha-3D_{\vartheta})}\Bigg((k_x^2-k_y^2) \cos(2\vartheta_0) + 2k_xk_y \sin(2\vartheta_0) \Bigg)\nonumber\\ 
    &\cdot \Bigg(\frac{\ex^{-4D_{\vartheta}t}-\ex^{-2\alpha t}}{2\alpha - 4D_{\vartheta}}-\frac{\ex^{-(\alpha + D_{\theta}t)}-\ex^{-2\alpha t}}{\alpha-D_{\vartheta}}\Bigg)\hat{p}^{(0)}(t,\vec{k}).
\end{eqnarray}
The anisotropic term, which is proportional to the initial
conditions, vanishes in the limit $t\rightarrow\infty$ for the same reasons as before
and 
also because the noise destroys the information of the initial condition.

Inspecting the structure of the perturbation series, one can convince oneself that all contributions 
of odd order are purely anisotropic and hence vanish in the long-time limit, while even orders
contain isotropic contributions independent from the initial conditions that prevail.

\subsubsection{The ``Donut'' Transition}
The problem how the probability distribution of an ABP in a trap
evolves from the trivial initial condition towards larger times
was studied e.g.~in \cite{Schehr19,PhysRevLett.129.158001}.
Let us consider a circular symmetric initial probability distribution around the origin 
e.g.~$P(\vec{q}_0) \sim \delta(\vec q_0)$. For vanishing or small activity 
(i.e.~small $v_0$), one expects diffusion to dominate and hence for long times, 
the distribution will spread out isotropically until the harmonic potential prevents further spread.
Increasing activity (i.e. speed $v_0$), at some point ballistic motion will dominate and 
is again limited by the potential.
Now rotational diffusion needs some time to decorrelate the direction of self-propulsion, 
which effectively yields an increased probability at a finite distance
from the center of the trap. One hence expects the probability to have a ring-shaped maximum
at finite distance and a minimimum at the origin.
This transition from a maximum at the center at small $v_0$ 
to a "Donut shape" at large $v_0$ can be discussed
using our perturbative calculation by looking at the curvature of the probability 
distribution at the origin in the long-time limit. 
We will restrict ourselves to this long-time limit here, but note that
one can also extract dynamic informations:
for instance, initially the distribution is anisotropic due to self-propulsion, cf.~Eq.~(\ref{pk1}) and 
higher orders, and only later the initial orientation gets decorrelated by the 
combined action of the trap and rotational diffusion, 
see e.g.~\cite{Schehr19,PhysRevLett.129.158001} for a visualization.

Choosing without loss of generality $P(\vec{q}_0) \sim \delta(\vec{q}_0)$ 
and $\vartheta_0=0$ and taking the limit $t\rightarrow\infty$, 
using Eqs.~(\ref{pk0}), (\ref{pk1}), (\ref{pk2})
we get the distribution up to second order as
\beq
    \hat{p}^{(2)}(\infty,\vec{k}) \equiv \hat{p}^{(2)}(\vec{k}) = \left(1-\frac{v_0^2\vec{k}^2}{4\alpha (\alpha + D_{\vartheta})}\right) \ex^{-\frac{D_q}{2\alpha}\vec{k}^2}.
\eeq
Performing the inverse Fourier transform using
standard Gaussian integral techniques, one finds
\beq\label{p_ord_2}
     p^{(2)}(\vec{x})=\left(\frac{\alpha^2v_0^2}{8\pi D_q^3(\alpha + D_{\vartheta})} \vec{x}^2+\frac{\alpha}{2\pi D_q}-\frac{\alpha v_0^2}{4\pi D_q^2(\alpha+D_{\vartheta})}\right) \ex^{-\frac{\alpha}{2D_q}\vec{x}^2}.
\eeq
It is easy to show that $\int_{\mathbb{R}^2}\mathrm{d}^2x\,p^{(2)}(\vec{x})=1$, i.e.~the
probability distribution is indeed normalized. Note that this is not obvious, because our 
perturbative approach in principle could have destroyed this property.

\begin{figure}[t!] 
    \centering
    \includegraphics[width=1.0\linewidth]{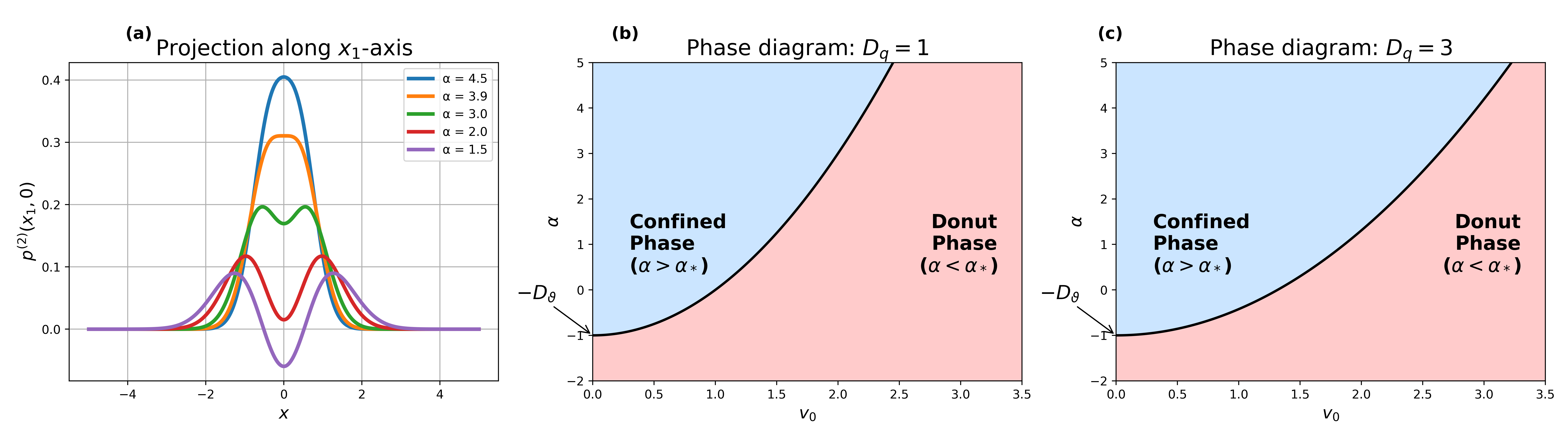}
    \caption{a) Shown is the long-time probability distribution up 
    to second order for different values of $\alpha$. 
    The distribution is isotropic and the $x$-direction was chosen for simplicity.
    Parameters are  $v_0=2$, $D_q=1$ and $D_{\vartheta}=0.1$, for varying potential 
    strengths $\alpha$. For these parameters, and to second order in the perturbation, 
    the ``Donut'' transition occurs at $\alpha_*=3.9$, cf.~Eq.~(\ref{crit_donut}).
    b), c) Shown are phase diagrams in the plane 
    potential strength $\alpha$ vs.~propulsion speed $v_0$ with $D_{\vartheta}=1$. 
    The line indicates the transition between the typical confined shape with a maximum 
    at the center and the ``donut'' shape. 
    Note that the curvature of the parabola decreases as $D_q$ increases, while
    its offset is given by $-D_{\vartheta}$.}
    \label{fig:Donut-Transition}
\end{figure}

Now, due to isotropy, it is sufficient to choose $\vec{x}=(x,0)$ and 
reduce the analysis to a one-dimensional function $p^{(2)}(x)$. 
The donut shape occurs, when the curvature at the origin becomes positive,
\beq
    \left.\frac{\mathrm{d}^2}{\mathrm{d} x^2}\right|_{x=0}p^{(2)}(x) = \frac{\alpha^2v_0^2}{2\pi D_q^3(\alpha+D_{\vartheta})} - \frac{\alpha^2}{2\pi D_q^2}>0,
\eeq
from which we can directly read off
\beq\label{crit_donut}
v_0^2>D_q(\alpha+D_\vartheta)\,\,\,{\rm or,\,equivalently}\,\,\,
\alpha<\alpha_*=\frac{v_0^2}{D_q}-D_\vartheta.
\eeq
It should be noted that the first result does not hold for $\alpha=0$, since there is no stationary
distribution in that case. Also, this result is not exact, since higher (even) orders in the expansion
still lead to corrections. 
Nevertheless, this result quantifies how, 
for the ``donut''-shaped distribution to be observable, 
the propulsion speed $v_0$ has to be sufficiently large to dominate over the diffusive processes in the potential.
Or alternatively, the potential has to be sufficiently weak 
in counteracting the propulsion, the latter being reduced by the decorrelation effect of rotational diffusion.
Fig.~\ref{fig:Donut-Transition}a) shows the distribution, Eq.~(\ref{p_ord_2}),
and b) and c) two phase diagrams, obtained using Eq.~(\ref{crit_donut}),
in the plane of potential strength $\alpha$ vs.~propulsion speed $v_0$.

\section{Brownian Circle Swimmer}
\label{secCircle}

Another interesting example to which our method can be directly applied  
is the Brownian circle swimmer (BCS) as studied in, e.g., Ref.~\cite{circle_simmer}. 
This kind of active particle has an 
additional constant angular velocity $\Omega_0$, such that the equations of motion are
\beq
    \derivative{t}{\vec q} &= v_0\vec n(\vartheta)-\alpha\vec q+\sqrt{2D_q}\,\vec\eta_q\,,\\
    \derivative{t}{\vartheta} &= \Omega_0 + \sqrt{2D_\vartheta}\,\eta_\vartheta\,.
\eeq
Physical causes for such an angular velocity are particle anisotropy 
for microswimmers \cite{Bechinger_circle}
or an off-axis internal driving force such as occurring for gliding microorganisms \cite{Leon_prep}. 

As for the ABP, we have to treat the self-propulsion perturbatively and 
the reference motion can be 
split into the translational and rotational dynamics. The translational reference motion and 
therefore the expression $Z_0^\alpha\left[\vec h_q,\vec g_q\right]$ does not change. 
In particular,  we want to stress here that we can easily include the case were the 
circle swimmer is trapped in an isotropic harmonic potential with stiffness $\alpha > 0$, 
which could be studied so far only in some limiting cases \cite{Loewen_trappedBCS}. 

\begin{figure}[t!]
    \centering
    \includegraphics[width=\linewidth]{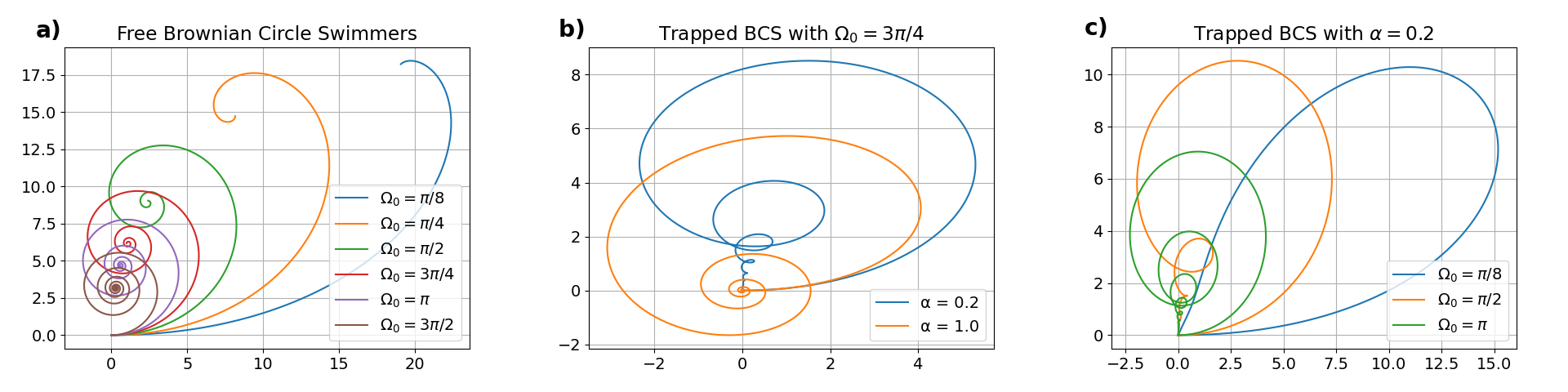}
    \caption{
    Mean positions $\mean{\vec q(t)}$ for a Brownian Circle swimmer (BCS) 
    starting at the origin and initially moving in the $x$-direction. 
    a) A free BCS for varying angular velocity $\Omega_0$. 
    b) A BCS with constant angular velocity $\Omega_0$  trapped in a harmonic potential  
    of varying stiffness $\alpha$. 
    c) Trapped BCS as in b) but now with constant harmonic trap stiffness $\alpha=0.2$ 
    and for varying angular velocity $\Omega_0$. 
    While the free BCS converges to a point that is determined by the parameters and the initial condition, 
    see Eq.~(\ref{freeBCS}),
    the trapped BCS always returns to the origin.}
    \label{fig:chiral-trajectories}
\end{figure}

The only change is the generating functional of the rotational reference motion,
where we have to incorporate the angular velocity. Thus, we have
\beq
  \hspace{-2cm}  Z_0^\vartheta[h_\vartheta,g_\vartheta] 
  &= \exp\left[\frac{1}{2}\left(h_\vartheta|C_\vartheta|h_\vartheta\right)+\left(h_\vartheta|G_\vartheta|g_\vartheta\right)\right]\times\nonumber\\&\quad\times
    \intd{\vartheta_0}\,P(\vartheta_0)\exp\left[\mathrm{i}\left(h_\vartheta|G_\vartheta|\vartheta_0\right)+\mathrm{i}\left(h_\vartheta|\Omega_0 t\right)\right],
\eeq
where without loss of generality we have again set the initial time to $t_0=0$. 
Following the calculations for the average position presented in the previous sections we find
\beq
    \hspace{-1cm}\mean{\vec q(t)}
    &= \mathrm{e}^{-\alpha t}\vec q_0 + \frac{v_0}{\left(\alpha-D_\vartheta\right)^2+\Omega_0^2}
    \bigg[\mathrm{e}^{-D_\vartheta t}\left[(\alpha-D_\vartheta)\vec n(t)-\Omega_0\vec n_\perp(t)\right]\nonumber\\
    &\hspace{5cm}-\mathrm{e}^{-\alpha t}\left[(\alpha-D_\vartheta)\vec n(0)-\Omega_0\vec n_\perp(0)\right]\bigg],
\eeq
where we have defined the direction vector $\vec n(t)=(\cos(\vartheta_0+\Omega_0 t),\sin(\vartheta_0+\Omega_0 t))^\intercal$ 
and  $\vec n_\perp(t)=(-\sin(\vartheta_0+\Omega_0 t),\cos(\vartheta_0+\Omega_0 t))^\intercal$, 
which is perpendicular to $\vec n$. 
For a vanishing potential ($\alpha=0$) we get the result for the free circle swimmer
\beq\label{freeBCS}
    \hspace{-1cm}\mean{\vec q(t)-\vec q_0} &=\frac{v_0}{D_\vartheta^2+\Omega_0^2}\left[D_\vartheta\vec n(0)+\Omega_0\vec n_\perp(0)-\mathrm{e}^{-D_\vartheta t}\left[D_\vartheta\vec n(t)+\Omega_0 \vec n_\perp(t)\right]\right],
\eeq
which was given in \cite{circle_simmer,Loewen}. 
The mean displacement of the free circle swimmer is, thus, 
a ``spira mirabilis'' that converges to a point, see Fig.~\ref{fig:chiral-trajectories}a),
given by the self-propulsion velocity $v_0$,
the rotational diffusion coefficient $D_\vartheta$, the angular velocity $\Omega_0$ 
and the initial direction angle $\vartheta_0$. 
On the other hand, the mean displacement for the trapped circle swimmer 
will always relax to the minimum of the potential, see Fig.~\ref{fig:chiral-trajectories}b), c).

It is also straightforward to calculate the MSD in the way presented before. For the free circle swimmer we get
\beq
    \hspace{-2.5cm}\mean{(\vec q(t)-\vec q_0)^2} &= 4D_qt+\frac{2v_0^2}{(D_\vartheta^2+\Omega_0^2)^2}\bigg[\Omega_0^2-D_\vartheta^2+D_\vartheta(D_\vartheta^2+\Omega_0^2)t\nonumber\\
    &\hspace{3cm} +\mathrm{e}^{-D_\vartheta t}\left[(D_\vartheta^2-\Omega_0^2)\cos(\Omega_0 t)-2D_\vartheta\Omega_0\sin(\Omega_0 t)\right]\bigg],
\eeq
which explicitly does not depend on the initial conditions 
for the position $\vec q_0$ and the direction angle $\vartheta_0$,
again in agreement with \cite{Loewen}.

\begin{figure}[t!] 
    \centering
    \includegraphics[width=\linewidth]{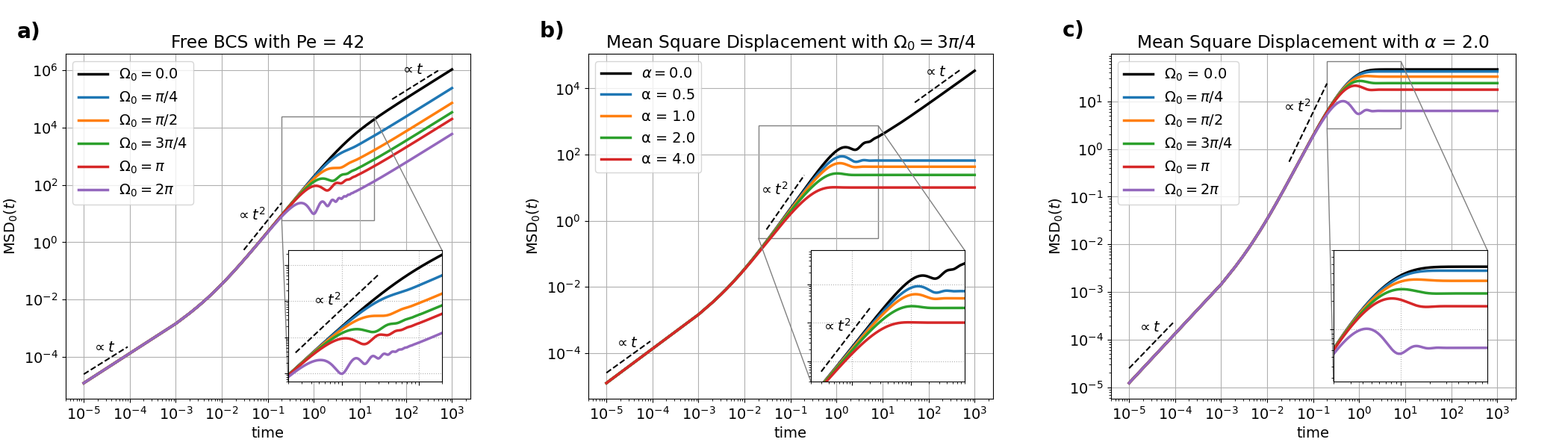}
    \caption{
    The mean square displacement for free and trapped Brownian Circle Swimmers. 
    Self-propulsion velocity $v_0$ and diffusion coefficients $(D_q,D_\vartheta)$ are 
    chosen as in Fig.~\ref{fig:MSD and D_eff ABP}. 
    a) The free BCS with varying angular velocities $\Omega_0$ shows oscillations 
    during the transition from the ballistic to the diffusive regime. The angular velocity 
    decreases the amplitude of the MSD at long timescales. 
    b) The trapped BCS for constant angular velocity $\Omega_0=3\pi/4~\mathrm{s}^{-1}$ 
    with varying stiffness $\alpha$ of the trap. 
    c) The trapped BCS for constant trap stiffness $\alpha=2.0$ and varying angular 
    velocity $\Omega_0$.}
    \label{fig:MSD and D_eff cABP}
\end{figure}

The expression for the general MSD of the trapped 
circle swimmer is a bit cumbersome to obtain and finally reads 
\beq
    \hspace{-2.5cm}\mean{(\vec q(t)-\vec q_0)^2} 
    &=\frac{2D_q}{\alpha}\left(1-\mathrm{e}^{-2\alpha t}\right)+\left(1-\mathrm{e}^{-\alpha t}\right)^2\mean{\vec q_0}\nonumber\\
    &\hspace{-0.5cm}+\frac{2\left(\mathrm{e}^{-\alpha t}-1\right)v_0}{\left(\alpha-D_\vartheta\right)^2+\Omega_0^2}\bigg[\mathrm{e}^{-D_\vartheta t}\left[(\alpha-D_\vartheta)\mean{\vec q_0\cdot\vec n(t)}-\Omega_0\mean{\vec q_0\cdot\vec n_\perp(t)}\right]\nonumber\\
    &\hspace{-0.5cm}\ \ \ \ \ \ \ \ \ \ \ \ \ \ \ \ \ \ \ \ \ \ \ \ \ \ -\mathrm{e}^{-\alpha t}\left[(\alpha-D_\vartheta)\mean{\vec q_0\cdot\vec n(0)}-\Omega_0\mean{\vec q_0\cdot\vec n_\perp(0)}\right]\bigg]\nonumber\\
    &\hspace{-0.5cm}+\frac{v_0^2}{\alpha\left((\alpha-D_\vartheta)^2+\Omega_0^2\right)\left((\alpha+D_\vartheta)^2+\Omega_0^2\right)}\bigg[\alpha\left(1+\ex^{-2\alpha t}\right)\left(\alpha^2-D_\vartheta^2+\Omega_0^2\right)\nonumber\\
    &\hspace{-0.5cm}\ \ \ \ \ \ \ \ \ \ \ \ \ \left.+D_\vartheta\left(1-\mathrm{e}^{-2\alpha t}\right)\left(-\alpha^2+D_\vartheta^2+\Omega_0^2\right)\right.\nonumber\\
    &\hspace{-0.5cm}\ \ \ \ \ \ \ \ \ \ \ \ \ \-2\alpha\mathrm{e}^{-(\alpha+D_\vartheta)t}\left[(\alpha^2-D_\vartheta^2+\Omega_0^2)\cos(\Omega_0 t)+2D_\vartheta\Omega_0\sin(\Omega_0 t)\right]\bigg].
\eeq
One can again see the same general structure as for the ABP:  
because the harmonic trap breaks homogeneity  
of space, 
the zeroth order depends on the initial position and the
first order correction depends on both initial position and initial orientation.
These contributions, however, decay for long times.
The MSD for both the free and trapped BCS are shown in 
Fig.~\ref{fig:MSD and D_eff cABP}
for the trivial initial condition. 
In the free case one observes 
that the angular velocity affects the transition from the ballistic to the diffusive regimes, 
i.e.~the behaviour for intermediate and late times. 
In particular, the angular velocity reduces the MSD for late times,
because the systematic rotation of the particle reverts the particle's direction 
of motion on those timescales. 
In the case of a trapped BCS we again observe a constant MSD at late times. 
The transition at intermediate times is determined by the trap stiffness $\alpha$. 
For weak traps, the transition happens later such that oscillations may still be visible. 
For strong traps the transition is very similar to the ABP,
with the angular velocity $\Omega_0$ reducing the late time MSD.

It is also possible to treat the probability distribution function
for the BCS, which we do not present in detail here. 
Since the translational reference motion is again a passive Brownian particle 
in an isotropic harmonic potential, the zero-th order probability distribution 
is still given by Eq.~(\ref{eq:prob-distr-0}), i.e.~the distribution of an Ornstein-Uhlenbeck process. 
The constant angular velocity of the BCS can affect the probability distribution 
only when the activity is taken into account, i.e.~starting at the first order 
in the perturbative expansion. 

Corrections are obtained that affect 
the probability distribution in the same way as for the non-chiral ABP. 
The linear order correction $\Delta^{(1)}\hat p(t,\vec k, \Omega_0)$ is purely 
anisotropic and corresponds to a ballistic shift of the distribution. This contribution 
is smaller than for the ABP due to $\Omega_0$ %>0$
and vanishes in the limit $t\rightarrow\infty$ due to the isotropy of the system. 
This is also in line with our results for the mean position of the particle. 

At quadratic order, the correction shows both anisotropic and isotropic contributions. 
In particular, the correction $\Delta^{(2)}\hat p(t,\vec k,\Omega_0)$ displays oscillations 
at frequency $\Omega_0$. In the long time limit, both, the anisotropic part and 
the oscillations become irrelevant and the probability distribution becomes isotropic. 
In comparison with the non-chiral ABP the width of the distribution is smaller, 
due to the circling of the particle with angular velocity $\Omega_0$. 
This behaviour is also visible in the MSD in Fig.~\ref{fig:MSD and D_eff cABP}. 
Depending on the time scales associated with rotational diffusion $D_\vartheta^{-1}$, 
the potential $\alpha^{-1}$ and the torque $\Omega_0^{-1}$, 
the transition from the ballistic (intermediate $t$) to the trapped regime (long-time $t$) 
displays oscillations. Furthermore, the long-time limit is smaller for larger angular 
velocities, which is in line with the narrower probability distribution for a BCS.

Importantly, the structure of the perturbative series does {\it not} change in comparison 
to the non-chiral ABP. This follows from the angular velocity changing the reference motion 
but {\it not} the activity operator. Thus, we can infer that also for the BCS corrections 
of odd order are purely anisotropic, vanishing in the long-time limit. 
In this limit only the isotropic corrections of even orders, that are independent of the initial
conditions, contribute to the probability distribution.

\section{Discussion and conclusions}
\label{secDisc}

Based on ideas put forward by Mazenko \cite{PhysRevE.81.061102}, 
we proposed a path-integral approach within the 
Martin-Siggia-Rose-Janssen-DeDominicis formalism
to treat the motion of active particles, including the presence of a harmonic trap. 
Importantly, within our formalism a passive particle in the harmonic potential
(equivalent to the Ornstein-Uhlenbeck process) can be solved completely analytically
and served as the reference state upon which the activity/self-propulsion 
is treated by systematic perturbation theory. 

We recovered known results for the ABP,
namely the mean position and the MSD could both be calculated exactly,
since for such moments/correlation functions the perturbation series
terminate at finite order. The probability distribution could also
be given as a series in propulsion speed $v_0$, with some insightful information
about the structure of the correction at successive orders 
(dependence on initial conditions, long-time limit). 
Our approach should be compared to the one in \cite{PhysRevLett.129.158001},
where a similar infinite series in terms of (complicated combinations of) 
generalized Laguerre polynomials was given.
Compared to that approach, relying on diagonalizing the 
Fokker-Planck/Smoluchowski operator using eigenfunctions,
which may be intractable for more complex systems (other potentials, more
complex motion involving additional active torques),
our approach seems to be more versatile:
if the reference motion can be solved -- which can
include harmonic potentials in both space and angle variable and, as shown in section \ref{secCircle}, 
a constant active torque --
the expansion stays the same. We hence could also treat the 
BCS, the Brownian circle swimmer experiencing an active torque, 
by a completely analogous calculation and obtain the general mean position and MSD in the presence of a trap.

It would be interesting -- and possible -- to generalize the approach to interacting
particles. We have shown here that the self-propulsion can be treated perturbatively
by applying the ``activity operator'' on the path integral/partition sum.
Particle interactions can be treated within the same framework by defining
an additional ``interaction operator'' involving the interaction potential.
We expect that the lowest order correlations for both hard or soft core repulsion
and for alignment interactions should be evaluable, allowing to extract information
about Motility-Induced Phase Separation (MIPS)  \cite{CatesTailleur} and the flocking transition
for Vicsek-type models \cite{Ginelli}, respectively, where the latter has been already 
discussed numerically in the presence of confinement in a harmonic trap \cite{RafaLuis}.

\bigskip

\noindent{\bf Data availability statement}\\
\noindent This work is analytical and all results are presented in detail.\\

\noindent{\bf Acknowledgments}\\
\noindent  We thank Matthias Bartelmann for stimulating discussions.
This work was funded by the Deutsche Forschungsgemeinschaft
(DFG, German Research Foundation) under Germany’s
Excellence Strategy EXC 2181/1 – 390900948 (the Heidelberg
STRUCTURES Excellence Cluster).

%\clearpage

\appendix

\section{Derivation of the Generating Functional}
\label{app1}

For a stochastic differential equation of first order as given by Eq.~(\ref{eq:stochastic-eom}),
\beq
    \mathrm{EoM}_\mu(t):=\derivative{t}{R_\mu} - F_\mu[R(t)] - \sigma_{\mu\nu}[R(t)]\eta_\nu(t)\equiv 0,\label{appendix:eq:stochastic-eom}
\eeq
with degrees of freedom $R_\mu(t)$, force $F_\mu[R(t)]$, diffusion  matrix $\sigma_{\mu\nu}[R(t)]$ 
and a Gaussian random process $\eta_\mu(t)$ as defined in Eq.~(\ref{eq:noisecorr}), 
one can formulate a path integral that allows, e.g., for the calculation of $n$-point correlation functions.
The derivation follows the calculations presented in \cite{PhysRevE.91.042103} closely. 
Note that here we will only consider additive noise, such that the diffusion matrix has constant entries only, 
i.e.\,$\sigma_{\mu\nu}[R(t)]=\mathrm{const}$.

A generating functional can be constructed from the path integral
\beq
    Z[h] = \int\mathcal{D}R\,P[R(t)]\exp\left(\intd{t}\,h(t)\cdot R(t)\right),
\eeq
where the functions $h_\mu(t)$ are external sources for the degrees of freedom. 
Each path $R(t)$ has to be weighted by its probability, which can be written as
\beq
    P[R(t)] = \int\mathcal{D}\eta\,\intd{R_0\,}~P\left[R(t)|R_0,\eta(t)\right]P(R_0)P[\eta(t)].
    \label{appendix:eq:total-probability}
\eeq
The total probability for a state $R(t)$ needs to be marginalised over the distribution $P(R_0)$ of initial conditions 
and the distribution 
\beq
    P[\eta(t)]=\exp\left(-\intd{t}\,\frac{\eta^2(t)}{2}\right)
\eeq
of the Gaussian random processes.
The conditional probability in Eq.~(\ref{appendix:eq:total-probability}) provides the dynamical constraint of 
the equations of motion, i.e.~only solutions $\underline{R}(t)$ to the stochastic differential equations can contribute to the path integral,
\beq
  \hspace{-1cm} P\left[R(t)|R_0,\eta(t)\right] &= \delta\left[R(t)-\underline R(t)\right]
    = \delta[\mathrm{EoM}_\mu(t)]\,\mathrm{det}\left[\funcderiv{R_\nu(t')}{\mathrm{EoM}_\mu(t)}\right],\label{Pcond}
\eeq
where we use the substitution rule for the Dirac delta distribution. We want to emphasize here that in contrast to a classical particle obeying Hamiltonian dynamics, this sharp transition probability is only valid for one particular realisation of the random processes $\eta(t)$. Thus, the total probability needs to involve the random processes.

In order to avoid solving any stochastic differential equation explicitly and eventually perform the integral over the random processes, we use a Fourier transform of the Dirac delta functional
\beq
    \delta[\mathrm{EoM}]=\int\mathcal{D}\chi\exp\left[-\mathrm{i}\intd{t}~\chi\cdot\left(\derivative{t}{R} - F[R(t)] - \sigma\eta(t)\right)\right],
\eeq
where we have introduced the auxiliary field $\chi_\mu(t)$ for each degree of freedom. In the Martin-Siggia-Rose \cite{PhysRevA.8.423} formalism $\chi_\mu(t)$ is referred to as the response field as it quantifies the response of the system to a delta-shaped perturbation.

The second term on the right of Eq.~(\ref{Pcond}) is the functional Jacobian, reading
\beq
    J_{\mu\nu}(t,t'):=\frac{\delta\mathrm{EoM}_\mu(t)}{\delta R_\nu(t')} = \left[\delta_{\mu\nu}\derivative{t}{}-\derivative{R_\nu(t')}{F_\mu[R(t)]}\right]\delta(t-t').\label{appendix:eq:Jacobian}
\eeq
Note that for multiplicative noise, there would be another contribution here from the derivative of the 
diffusion matrix and the random processes. In the case of additive noise considered here, however,  
there is no dependence on $\eta(t)$ in the functional Jacobian. In $Z[h]$, 
we can hence perform the integral over the random processes, which reads
\beq
  \hspace{-1.5cm}  \mathcal{I}_\eta &= \int\mathcal{D}\eta\,\exp\left(-\intd{t}\,\left[\frac{\eta^2(t)}{2}+\mathrm{i}\chi\cdot\sigma\eta(t)\right]\right)
    = \exp\left(-\frac{1}{2}\intd{t}\,\chi\cdot\sigma^2\chi\right).
\eeq
In two dimensions, the correlation matrix is simply given by $\sigma^2=2k_\mathrm{B}T\mathrm{diag}(D_x,D_y)$ 
(and analogously in higher dimensions).

Now we have to treat the
determinant of the Jacobian (\ref{appendix:eq:Jacobian}). 
By a functional integration over a set of conjugate Grassmann functions 
$\{\bar\omega_\mu(t),\omega_\mu(t)\}$ it can be written  in the following way
\beq
    \hspace{-1cm}\mathrm{det}\left[J_{\mu\nu}(t,t')\right] = \int\mathcal{D}\bar\omega\int\mathcal{D}\omega\,\exp\left(\intd{t}\,\intd{t}'\,\,\bar\omega_\mu(t) J_{\mu\nu}(t,t')\omega_\nu(t')\right).
\eeq
This is a consequence of the anticommuting property of Grassmann functions and the fact that integration and differentiation are identical operations in a Grassmann algebra, cf.~\cite{Zinn-Justin}. Since we are not interested in the statistics of these Grassmann functions, we will integrate them out immediately. For that purpose, we need to solve an integral of the form
\beq
    \mathcal{I}_\omega &= \int\mathcal{D}\bar\omega\int\mathcal{D}\omega\,\exp\left[\intd{t}\,\bar\omega\cdot\frac{\mathrm{d}}{\mathrm{d}t\,}\omega\right]\exp\left[A(\bar\omega,\omega)\right],\label{appendix:eq:Grassmann-integral}
\eeq
with the function
\beq
    A(\bar\omega,\omega) = -\intd{t}\,\bar\omega_\mu(t)\derivative{R_\nu(t)}{F_\mu[R(t)]}\omega_\nu(t).
\eeq
The integral $\mathcal{I}_\omega$ is a Gaussian integral, which may be solved by Wick's theorem. We define the differential operator $\hat D$ and its inverse $G_\mathrm{R}$ as
\beq
    \hat D = \frac{\mathrm{d}}{\mathrm{d}t\,},\quad G_\mathrm{R}(t,t')=\Theta(t-t'),\quad \hat DG_\mathrm{R}(t,t')=\delta(t-t').
\eeq
The function $G_\mathrm{R}$ is nothing but 
the Heaviside step function, i.e.~the retarded Green's function of the time derivative, 
since we want a causal theory. Now that the correlator of the Grassmann sector is known, 
we are able to evaluate the integral. It is the Gaussian expectation value of the second exponential function. 
Wick's theorem for Grassmann functions explicitly says
\beq
    &\left<\bar\omega_{\mu_1}(t_1)\omega_{\nu_1}(t'_1)\dots\bar\omega_{\mu_n}(t_n)\omega_{\nu_n}(t'_n)\right> \nonumber\\ &\quad = \sum_{\mathrm{permutations\, P}}\mathrm{sgn}(\mathrm{P})\left<\bar\omega_{\mu_1}(t_1)\omega_{\nu_{\mathrm{P}_1}}(t'_{\mathrm{P}_1})\right>\dots\left<\bar\omega_{\mu_n}(t_n)\omega_{\nu_{\mathrm{P}_n}}(t'_{\mathrm{P}_n})\right>\label{appendix:eq:Wick}
\eeq
with the expectation values given by
\beq
    \hspace{-1cm}\left<\bar\omega_\mu(t)\omega_\nu(t')\right>=\delta_{\mu\nu}G_\mathrm{R}(t,t'),\quad\mathrm{and}\quad\left<\bar\omega_\mu(t)\bar\omega_\nu(t')\right> =0=\left<\omega_\mu(t)\omega_\nu(t')\right>.\label{appendix:eq:Grassmann-expectation}
\eeq
Expanding the second exponential of (\ref{appendix:eq:Grassmann-integral}) in a Taylor series we find 
\beq
    \hspace{-2.5cm}\exp\left[A(\bar\omega,\omega)\right]&=1-\intd{t}\,\derivative{R_\nu(t)}{F_\mu[R(t)]}\bar\omega_\mu(t) \omega_\nu(t)\nonumber\\ &+\frac{1}{2!}\intd{t}\,\intd{t}'\,\derivative{R_\nu(t)}{F_\mu[R(t)]}\derivative{R_\lambda(t')}{F_\gamma[R(t')]}\bar\omega_\mu(t) \omega_\nu(t)\bar\omega_\gamma(t') \omega_\lambda(t')-\dots
\eeq
We can now use Wick's theorem and (\ref{appendix:eq:Grassmann-expectation}) to evaluate the Grassmann integral in terms of 2-point functions. First we notice that any term involving expectation values of the same (conjugate) function vanishes regardless of component. Secondly, we have a causal theory, such that
\beq
    \left<\bar\omega_\mu(t) \omega_\lambda(t')\right>\left<\bar\omega_\kappa(t') \omega_\nu(t)\right>=\delta_{\mu\nu}\delta_{\kappa\lambda}G_\mathrm{R}(t,t')G_\mathrm{R}(t',t)\equiv 0
\eeq
except for $t=t'$. Thus, the integral is eventually evaluated to be
\beq
    \mathcal{I}_\omega &= \exp\left[-G_\mathrm{R}(0)\intd{t}\,\derivative{R_\mu(t)}{F_\mu[R(t)]}\right]
\eeq
with a sum over repeated indices. The value of the Heaviside theta function $G_\mathrm{R}(0)$ evaluated 
at zero depends on the stochastic prescription. In case of a retarded or It\^o prescription it is zero, 
while for the symmetric Stratonovich prescription one chooses $G_\mathrm{R}(0)=1/2$. 
The advantage of the Stratonovich prescription is that it leaves all calculus unchanged, 
while in the It\^o prescription a drift term in the derivatives needs to be introduced as stated by It\^o's lemma
\cite{Oksendal}.

Collecting all terms and adding the second source field $g_\mu(t)$,
we arrive at the generating functional for a first order stochastic differential equation 
as given in Eqs. (\ref{eq:Z[h,g]})-(\ref{eq:action-general}), namely,
\beq
    \hspace{-2.5cm}Z[h,g] &= \int\mathcal{D}R\int\mathcal{D}\chi\intd{R_0}\,P(R_0)\exp\left(-S[R,\chi]\right)\exp\left(\intd{t}\,\left[h\cdot R+g\cdot\chi\right]\right),\label{appendix:eq:generating-functional}
\eeq
with the action
\beq
    S[R,\chi] &= \intd{t}\,\left[\frac{1}{2}\chi_\mu\sigma_{\mu\gamma}\sigma_{\nu\gamma}\chi_\nu+\mathrm{i}\chi_\mu\left(\frac{\mathrm{d}R_\mu}{\mathrm{d}t\,}-F_\mu\right)+G_\mathrm{R}(0)\frac{\mathrm{d}F_\mu}{\mathrm{d}R_\mu}\right].\label{appendix:eq:action-general}
\eeq
This result was used in \cite{PhysRevE.81.061102} for the description of a particle ensemble obeying Smoluchowski dynamics. 
A discussion on the relation of the Grassmann functions and the stochastic prescription can be found 
in \cite{Barci}. A derivation of the generating functional including multiplicative noise is given 
in \cite{PhysRevE.91.042103}, which we followed for our presentation.

\section{Integrals for the second order}\label{app3}

Here we present some details on the calculation of the expressions
\beq
    (\sigma\varepsilon):=\int_0^\infty \mathrm{d}t' \int_0^\infty \mathrm{d}t'' G(t,t')G(t,t'')Z^{\vartheta}_{\sigma\varepsilon}[-\mathrm{i}\sigma\delta_{t'}-\mathrm{i}\varepsilon\delta_{t''},0],
\eeq
where $\sigma,\varepsilon\in\{+,-\}$. 
We begin with inserting the propagators $C_{\vartheta}$ in $Z^{\vartheta}_0$, which yields
\beq
    Z^{\vartheta}_{0}[-\mathrm{i}\sigma\delta_{t'}-\mathrm{i}\varepsilon\delta_{t''}]
    &=\ex^{-\mathrm{i}(\sigma+\varepsilon)\vartheta_0-\frac{1}{2}C_{\vartheta}(t',t')-\frac{1}{2}C_{\vartheta}(t'',t'')-\sigma\varepsilon C_{\vartheta}(t',t'')}\nonumber\\
    &=\ex^{-\mathrm{i}(\sigma+\varepsilon) \vartheta_0-D_{\vartheta}t'-D_{\vartheta}t''-\sigma \varepsilon[\theta(t'-t'')t''+\theta(t''-t')t']}.
\eeq
We will start with the easiest one to calculate, which is $(+-)=(-+)$ where the initial condition cancels out.
One finds 
\beq % 
 \hspace{-2.cm}
    (+-)&
 \hspace{-1.cm}    
    = i^2\int_{t'}\int_{t''}\theta(t-t')\theta(t-t'')\ex^{\alpha(t'-t)}\ex^{\alpha(t''-t)}\ex^{-D_{\vartheta}t'-D_{\vartheta}t'' + 2D_{\vartheta}[\theta(t''-t')t'+\theta(t'-t'')t'']}\nonumber\\
    &  \hspace{-1.cm} 
    =-\int_0^{t}\mathrm{d} t'\int_0^{t}\mathrm{d} t''\ex^{\alpha(t'+t''-2t)}\ex^{-D_{\vartheta}\big[t'(1-2\theta(t''-t'))+t''(1-2\theta(t'-t''))\big]}\nonumber\\
    &  \hspace{-1.cm} 
    =- \int_0^{t}\mathrm{d} t'\int_0^{t}\mathrm{d} t'' \ex^{\alpha(t'+t''-2t)}\ex^{-D_{\vartheta}|t'-t''|}.
\eeq
The last integral can be calculated by splitting the integration domain $[0,t]\times[0,t]$. 
Since the argument is invariant under the reversal $t'\rightarrow t''$ and vice versa, 
it is enough to calculate the integral once and multiply it by $2$, yielding
\beq
    (+-)&=-2\ex^{-2\alpha t}\int_0^{t}\mathrm{d} t'\int_0^{t}\mathrm{d} t'' \ex^{\alpha(t'+t'')-D_{\vartheta}(t'-t'')}\nonumber\\
%        &=-2\ex^{-2\alpha t}\Bigg(\int_0^{t}\mathrm{d} t' \ \ex^{t'(\alpha-D_{\vartheta})}\Bigg)\Bigg(\int_0^{t'}\mathrm{d} t'' \ \ex^{t''(\alpha+D_{\vartheta})}\Bigg)\nonumber\\
%        &=-2\ex^{-2\alpha t}\int_0^{t}\mathrm{d} t' \ \ex^{t'(\alpha-D_{\vartheta})} \Bigg(\frac{1}{\alpha+D_{\vartheta}}\Bigg)\Bigg(\ex^{t'(\alpha+D_{\vartheta})}-1\Bigg)\nonumber\\
        &=-\frac{2}{\alpha+D_{\vartheta}}\Bigg(\frac{1-\ex^{-2\alpha t}}{2\alpha}-\frac{\ex^{-t(\alpha+D_{\vartheta})}-\ex^{-2\alpha t}}{\alpha-D_{\vartheta}}\Bigg).
\eeq

Next we calculate the combinations $(++)$ and $(--)$. One can easily see that they differ only by the contribution 
of the initial condition $\vartheta_0$ and therefore the integral is again symmetric under the exchange of $t'$ and $t''$. 
In the same way as above we can manipulate the integral expressions and after introducing the factor of $2$ from 
the symmetric time ordering, we can evaluate 
\beq
    (\sigma\sigma)&=-2\int_0^t\mathrm{d} t'\int_0^t\mathrm{d} t'' \ \ex^{\alpha(t'+t'')-2\alpha t - 2i\sigma\vartheta_0-D_{\vartheta}(3t''+t')}\nonumber\\
    &=-2\ex^{-2i\sigma\vartheta_0-2\alpha t}\int_0^t\mathrm{d} t' \ex^{(\alpha -D_{\vartheta}t')}\int_0^{t'}\mathrm{d} t'' \ \ex^{(\alpha-3D_{\vartheta})t''}\nonumber\\
%    &=-2\ex^{-2i\sigma\vartheta_0-2\alpha t}\int_0^{t}\mathrm{d} t' \frac{\ex^{(\alpha-D_{\vartheta})t'}}{\alpha-3D_{\vartheta}}\Bigg(\ex^{(\alpha-3D_{\vartheta})t'}-1\Bigg)\nonumber\\
%    &=-\frac{-2\ex^{-2i\sigma\vartheta_0-2\alpha t}}{\alpha-3D_{\vartheta}}\int_0^{t}\mathrm{d} t' \Bigg(\ex^{(2\alpha t-4D_{\vartheta})t'}-\ex^{(\alpha-D_{\vartheta}t')}\Bigg)\nonumber\\
    &=-\frac{-2\ex^{-2i\sigma\vartheta_0}}{\alpha-3D_{\vartheta}}\Bigg(\frac{\ex^{-4\alpha D_{\vartheta}t}-\ex^{-2\alpha t}}{2\alpha-4D_{\vartheta}}-\frac{\ex^{-(\alpha+D_{\vartheta})t'}-\ex^{-2\alpha t}}{\alpha-D_{\vartheta}}\Bigg).
\eeq
With $(+-)=(-+)$,  $(++)$ and $(--)$ we have all we need to calculate the expressions for the MSD 
and the probability distribution in second order $\mathcal{O}(v_0^2)$ as described in the main text.

\section*{References}
\bibliographystyle{unsrt.bst}\bibliography{ABP}

\begin{thebibliography}{10}

\bibitem{Romanczuk}
P.~Romanczuk, M.~Bär, W.~Ebeling, B.~Lindner, and L.~Schimansky-Geier.
\newblock Active {Brownian} particles.
\newblock {\em Eur. Phys. J. Special Topics}, 202:1, 2012.

\bibitem{Bergbook}
H.~C. Berg.
\newblock {\em Random Walks in Biology}.
\newblock Princeton University Press, 1993.

\bibitem{WadhwaBerg}
N.~Wadhwa and H.~C. Berg.
\newblock Bacterial motility: Machinery and mechanisms.
\newblock {\em Nat. Rev. Microbiol.}, 20:161, 2022.

\bibitem{BetaStark}
M.~Seyrich, Z.~Alirezaeizanjani, C.~Beta, and H.~Stark.
\newblock Statistical parameter inference of bacterial swimming strategies.
\newblock {\em New J. Phys.}, 20:103033, 2018.

\bibitem{FletcherTheriot}
D.~A. Fletcher and J.~A. Theriot.
\newblock An introduction to cell motility for the physical scientist.
\newblock {\em Phys. Biol.}, 1:T1, 2004.

\bibitem{Hakim_review}
V.~Hakim and P.~Silberzan.
\newblock Collective cell migration: a physics perspective.
\newblock {\em Rep. Prog. Phys.}, 80:076601, 2017.

\bibitem{Leon_prep}
L.~Lettermann, F.~Ziebert, M.~Singer, F.~Frischknecht, and U.~S. Schwarz.
\newblock Three-dimensional chiral active {Ornstein-Uhlenbeck} model for
  helical motion of microorganisms.
\newblock {\em Phys. Rev. Lett.}, 135:128403, 2025.

\bibitem{Howse_Ramin07}
J.~R. Howse, R.~A.~L. Jones, A.~J. Ryan, T.~Gough, R.~Vafabakhsh, and
  R.~Golestanian.
\newblock Self-motile colloidal particles: From directed propulsion to random
  walk.
\newblock {\em Phys. Rev. Lett.}, 99:048102, 2007.

\bibitem{BartoloQuincke}
A.~Bricard, J.~B. Caussin, N.~Desreumaux, O.~Dauchot, and D.~Bartolo.
\newblock Emergence of macroscopic directed motion in populations of motile
  colloids.
\newblock {\em Nature}, 503:95, 2013.

\bibitem{IgorMagnetic}
A.~Kaiser, A.~Snezhko, and I.~S. Aranson.
\newblock Flocking ferromagnetic colloids.
\newblock {\em Sci. Adv.}, 3:e1601469, 2017.

\bibitem{RevModPhys.88.045006}
C.~Bechinger, R.~Di~Leonardo, H.~L\"owen, C.~Reichhardt, G.~Volpe, and
  G.~Volpe.
\newblock Active particles in complex and crowded environments.
\newblock {\em Rev. Mod. Phys.}, 88:045006, Nov 2016.

\bibitem{Szamel}
G.~Szamel.
\newblock Self-propelled particle in an external potential: Existence of an
  effective temperature.
\newblock {\em Phys. Rev. E}, 90:012111, 2014.

\bibitem{Bechinger_trap}
S.~Jahanshahi, C.~Lozano, B.~Liebchen, H.~Löwen, and C.~Bechinger.
\newblock Realization of a motility-trap for active particles.
\newblock {\em Commun. Phys.}, 3:127, 2020.

\bibitem{exp_multistable}
A.~Militaru, M.~Innerbichler, M.~Frimmer, F.~Tebbenjohanns, L.~Novotny, and
  C.~Dellago.
\newblock Escape dynamics of active particles in multistable potentials.
\newblock {\em Nat Commun.}, 12:2446, 2021.

\bibitem{circle_simmer}
S.~van Teeffelen and H.~Löwen.
\newblock Dynamics of a {Brownian} circle swimmer.
\newblock {\em Phys. Rev. E}, 78:020101, 2008.

\bibitem{Bechinger_circle}
F.~Kümmel, B.~{ten Hagen}, R.~Wittkowski, I.~Buttinoni, R.~Eichhorn, G.~Volpe,
  H.~Löwen, and C.~Bechinger.
\newblock Circular motion of asymmetric self-propelling particles.
\newblock {\em Phys. Rev. Lett.}, 110:198302, 2013.

\bibitem{circle_swim_newer}
A.~Kirvin, D.~Gregory, A.~Parnell, A.~I. Campbell, and S.~Ebbens.
\newblock Rotating ellipsoidal catalytic micro-swimmers via glancing angle
  evaporation.
\newblock {\em Mater. Adv.}, 2:7045, 2021.

\bibitem{Loewen}
H.~Löwen.
\newblock Inertial effects of self-propelled particles: From active {Brownian}
  to active {Langevin} motion.
\newblock {\em J. Chem. Phys.}, 152(4):040901, 01 2020.

\bibitem{Vicsek}
T.~Vicsek, A.~Czirók, E.~{Ben-Jacob}, I.~Cohen, and O.~Shochet.
\newblock Novel type of phase transition in a system of self-driven particles.
\newblock {\em Phys. Rev. Lett.}, 75:1226, 1995.

\bibitem{Ginelli}
F.~Ginelli.
\newblock The physics of the {Vicsek} model.
\newblock {\em Eur. Phys. J. Special Topics}, 225:2099, 2016.

\bibitem{Schehr19}
U.~Basu, S.~N. Majumdar, A.~Rosso, and G.~Schehr.
\newblock Long-time position distribution of an active {Brownian} particle in
  two dimensions.
\newblock {\em Phys. Rev. E}, 100:062116, 2019.

\bibitem{Malakar20}
K.~Malakar, A.~Das, A.~Kundu, K.~V. Kumar, and A.~Dhar.
\newblock Steady state of an active {Brownian} particle in a two-dimensional
  harmonic trap.
\newblock {\em Phys. Rev. E}, 101:022610, 2020.

\bibitem{PhysRevLett.129.158001}
M.~Caraglio and T.~Franosch.
\newblock Analytic solution of an active {Brownian} particle in a harmonic
  well.
\newblock {\em Phys. Rev. Lett.}, 129:158001, 2022.

\bibitem{Pruessner24}
Z.~Zhang, L.~Fehertoi-Nagy, M.~Polackova, and G.~Pruessner.
\newblock Field theory of active {Brownian} particles in potentials.
\newblock {\em New J. Phys.}, 26:013040, 2024.

\bibitem{RafaLuis}
R.~González-Albaladejo, A.~Carpio, and L.~L. Bonilla.
\newblock Scale-free chaos in the confined {Vicsek} flocking model.
\newblock {\em Phys. Rev. E}, 107:014209, 2023.

\bibitem{PhysRevE.81.061102}
G.~F. Mazenko.
\newblock Fundamental theory of statistical particle dynamics.
\newblock {\em Phys. Rev. E}, 81:061102, 2010.

\bibitem{PhysRevE.91.042103}
M.~V. Moreno, Z.~G. Arenas, and D.~G. Barci.
\newblock Langevin dynamics for vector variables driven by multiplicative white
  noise: A functional formalism.
\newblock {\em Phys. Rev. E}, 91:042103, 2015.

\bibitem{PhysRevE.76.011123}
A.~W.~C. Lau and T.~C. Lubensky.
\newblock State-dependent diffusion: Thermodynamic consistency and its path
  integral formulation.
\newblock {\em Phys. Rev. E}, 76:011123, 2007.

\bibitem{DasMazenko}
S.~P. Das and G.~F. Mazenko.
\newblock Field theoretic formulation of kinetic theory: Basic development.
\newblock {\em J. Stat. Phys.}, 149, 2012.

\bibitem{PhysRevA.8.423}
P.~C. Martin, E.~D. Siggia, and H.~A. Rose.
\newblock Statistical dynamics of classical systems.
\newblock {\em Phys. Rev. A}, 8:423, 1973.

\bibitem{PhysRevD.40.3363}
E.~Gozzi, M.~Reuter, and W.~D. Thacker.
\newblock Hidden {BRS} invariance in classical mechanics. {II}.
\newblock {\em Phys. Rev. D}, 40:3363--3377, 1989.

\bibitem{Barci}
Z.~G. Arenas and D.~G. Barci.
\newblock Functional integral approach for multiplicative stochastic processes.
\newblock {\em Phys. Rev. E}, 81:051113, 2010.

\bibitem{Risken}
H.~Risken.
\newblock {\em The Fokker-Planck Equation}.
\newblock Springer, 1989.

\bibitem{Loewen_trappedBCS}
S.~Jahanshahi, H.~Löwen, and B.~ten Hagen.
\newblock Brownian motion of a circle swimmer in a harmonic trap.
\newblock {\em Phys. Rev. E}, 95:022606, 2017.

\bibitem{CatesTailleur}
M.~E. Cates and J.~Tailleur.
\newblock Motility-induced phase separation.
\newblock {\em Annu. Rev. Condens. Matter Phys.}, 6:219, 2015.

\bibitem{Zinn-Justin}
J.~Zinn-Justin.
\newblock {\em Quantum Field Theory and Critical Phenomena}.
\newblock Oxford University Press, 2021.

\bibitem{Oksendal}
B.~\O ksendal.
\newblock {\em Stochastic Differential Equations: An Introduction with
  Applications}.
\newblock Springer Berlin, Heidelberg, 2003.

\end{thebibliography}

\end{document}